\documentclass[12pt,preprint]{aastex}
\usepackage{amsmath}

\shorttitle{Formation and Evolution of CO White Dwarfs}
\shortauthors{Ablimit at el.}

\begin{document}


\title{The CO White Dwarf $+$ Intermediate Mass/Massive Star Binary Evolution: Possible Merger Origins for Peculiar Type Ia and II Supernovae}

\author{Iminhaji Ablimit\altaffilmark{1,2}}

\altaffiltext{1}{Key Laboratory for Optical Astronomy, National Astronomical Observatories, Chinese Academy of Sciences, Beijing 100012,
China. iminhaji@nao.cas.cn}
\altaffiltext{2}{Department of Astronomy, Kyoto University, Kitashirakawa-Oiwake-cho, Sakyo-ku, Kyoto 606-8502, Japan.}




\begin{abstract}

Binary stellar evolution has been studied as important pathway to
initiate various transient events like supernovae (SNe).
Although the common envelope (CE) in a binary, outcomes
of the CE and conditions for the SN explosion during the CE phase
are uncertain, it has been suggested that SN explosions can be triggered
during the CE phase. In this work, we explore formation and evolution
routes of carbon/oxygen (CO) WDs binaries in order to investigate mergers of CO
WDs and cores of non-degenerate stars during the CE phase as possible
origins for SNe under the core merger
detonation (CMD) scenario by considering several binary physical models.
Evolution of CO WD + intermediate mass normal (non-degenerate and hydrogen-rich) star binaries lead to
mergers during the CE phases still may trigger type Ia SNe (SNe Ia) interacting with circumstellar
material under different models. Mergers
between CO WDs and cores of He (non-degenerate and helium-rich) stars within the CE phases are rare comparing to other CE merger events.
These two channels may produce peculiar SNe Ia such as over-luminous/super-chandrasekhar mass SNe Ia as certain fraction of
them have combined core masses $\geq2M_\odot$. In the channel of CO WD +
massive star ($\geq8M_\odot$) CE events, we find that rates of mergers
between CO WDs and He cores of massive stars are (0.85 -
12.18)$\times10^{-5}{M_\odot}^{-1}$ which may initiate type II superluminous SNe like SN 2006gy, and delay
times from this scenario are in the range of 34 - 120 Myr.
Our results based on the CMD model are comparable with observational results of peculiar SNe.

\end{abstract}

\keywords{Unified Astronomy Thesaurus concepts: Binary stars (154); Supernovae (1668); Close binary stars (254);
X-ray binary stars (1811);  Common envelope evolution (2154); Stellar evolution (1599);
White dwarf stars (1799); Neutron stars (1108); Gravitational waves (678);
}

\section{Introduction}

The explosion of a carbon/oxygen white dwarf (CO WD)
has been commonly accepted for generating a type Ia supernova (SN Ia). In
the main progenitor scenarios of SNe Ia, CO WDs accrete matter from
their non-degenerate companions until its mass reaches the
Chandrasekhar limit mass or merges with other degenerate ones to
explode as SNe Ia (Tutukov \& Yungelson 1979; Nomoto 1982; Iben \& Tutukov 1984;
Yungelson et al. 1994; Li \& van den Heuvel 1997; Langer et al. 2000; Han \& Podsiadlowski 2004;
Hachisu et al. 2008; L$\ddot{\rm u}$ et al. 2009; Ruiter et al. 2011;
Toonen et al. 2012; Hamers et al. 2013; Ablimit et al. 2016; Wang et al. 2017; Livio \&
Mazzali 2018). The problem of the evolutionary scenarios
that may lead to SNe Ia still remains uncertain. It is still not
known, whether any of the suggested single-degenerate, double-degenerate,
core-degenerate etc. channels may lead to explosion, whether one or several
channels may contribute.
In the past decades, many dozens of
works have been attempting to solve the problem by using binary population synthesis
(BPS) method (recent reviews, e.g., Maeda \& Terada 2016; Livio \&
Mazzali 2018).

Binaries should
survive from the CE (Paczy$\acute{\rm
n}$ski 1976) phase in those scenarios if they evolve into the CE. In
addition to the two main scenarios, Sparks \& Stecher (1974) first
proposed that the SN explosion may happen by the merger inside the
CE, and this indicates that the merger inside the CE might be a
possible trigger for various SNe (not just for SNe Ia). There is a
type of merger model during the CE phase, the so-called
core-degenerate scenario which is a merger of a WD with the hot CO
core of the asymptotic giant branch (AGB) star during
the CE evolution that may trigger an SN Ia, and there would be a
circumstellar material (CSM) interaction in this model
(e.g., Livio \& Riess 2003; Kashi \& Soker
2011; Soker 2011; Soker et al. 2013; Soker 2019). Although the rate
of the core degenerate model is controversial (Wang et al. 2017; Canals et al. 2018),
this model has
been suggested as the possible progenitors for peculiar SNe Ia-CSM, i.e., SN 2014J and  PTF
11kx where the large CSM mass results from the evolution of a CO WD + intermediate mass star
($<8M_\odot$) system (e.g., Soker 2011; Soker et al. 2013; Soker
2019). One typical observational result for supporting the merger of a WD with the AGB star
core process might be the
 optical observations of SN 2002ic (Hamuy et al. 2003) which
reveal large amounts of CSM seen as a strong hydrogen emission. Hamuy et al. (2003)
suggested that the progenitor system contained a massive AGB
star which lost a few solar masses of hydrogen-rich gas prior to the type Ia
explosion.
Briggs et al. (2015) performed the BPS calculation for the merger of CO WDs +
AGB cores, and argued that the majority of CO WDs with strong
magnetic fields might be resulted from mergers during the CE phase. It is also worth to note that the recent
detailed CO WD binary evolution study by Ablimit \& Maeda (2019a)
showed that the strong magnetic field of a CO WD may play a key role
in a different way during the mass transfer, so that highly
magnetized CO WD binaries may have significant contributions to SNe
Ia without merging inside the CE (see also Ablimit \& Maeda 2019b).
Considering various SN Ia populations including SNe Ia-CSM, we will investigate
mergers during the CE phase as the possible origins for SNe Ia.

Unlike SNe Ia, core-collapse SNe (CCSNe)
originate from massive stars. The binary evolution including mass loss
and/or merger has influence stellar populations and progenitors of
type II CCSNe (e.g., Eldridge et al. 2008; Zapartas et
al.2017). Super-luminous supernovae (SLSNe) are type of stellar explosions with
the luminosity 10 or more times higher than that of standard supernovae,
and several mechanisms are proposed to generate SLSNe including their light-curves and spectral features (e.g., Gal-Yam 2019).
Type I SLSNe which is one of the two main classes of SNe
do not show hydrogen lines in their spectra (e.g., Pastorello et al.
2010; Inserra et al. 2013; Quimby et al. 2011), while the other main class type II
SLSNe exhibit hydrogen lines (e.g., Smith et al. 2007; Ofek et al.
2007) and result from massive stars (e.g., Woosely et al. 2007; Moriya et al. 2013). Sabach
\& Soker (2014) pointed out that the CE merger-induced events may be
the origins of SNe-like transients. Recently, Jerkstrand et al.
(2020) presented their observational and theoretical results for a
type II SLSN named SN 2006gy (e.g., Smith et al. 2007; Ofek et al.
2007; Agnoletto et al. 2009), and demonstrated that the observed
signature in the late-time nebular spectrum and light curve of this
transient can be reproduced by the iron-rich SN Ia ejecta and its
interaction with a massive ($\sim 10M_\odot$) CSM.
Jerkstrand et al. (2020) suggested to explain SN 2006gy with one evolution pathway proposed by Sabach
\& Soker (2014) which is that of a
merger of a CO WD with the core of a red supergiant (RSG) star.
Based on Sabach \& Soker (2014) and
Jerkstrand et al. (2020),
we will study the possible progenitor model for SN 2006gy-like SLSNe under a new model (see section 2 for the model description)
with different binary physics by considering several important evolution pathways.

After a WD $+$ a massive non-degenerate star formation
from a primordial MS-MS binary, the massive star may fill its
Roche lobe, undergo Roche-lobe overflow (RLOF) and start the mass transfer.
Then, the WD can be engulfed by the companion's envelope due to the
rapid and unstable mass transfer (Paczy$\acute{\rm n}$ski 1976).
Then, the WD may merge with the massive companion's core during the
CE phase, or the binary may expel the CE and eventually may become a
neutron star (NS) - WD system which is one of the promising
gravitational wave (GW) and gamma ray burst sources (e.g., van Paradijs et al. 2000). Previous works
mainly studied the formation of NS - WD systems that formed from binary evolutions (e.g., Tutukov \& Yunggelson 1993; Tauris \& Sennels
2000; O'Shaughnessy \& Kim 2010; Sabach \& Soker 2014). In this
work, we study the formation and evolution of a CO WD with a massive
star ($\geq 8 M_\odot$) by considering survival from or merger
inside the CE. A CO WD- NS will be formed if they survive from the
CE, or may produce SN 2006gy-like transients if they merge inside
the CE.

By adopting BPS models including a wide range of physical ingredients,
we investigate evolutionary processes in CO WD binaries by mainly focusing on the possible
merger-induced detonation inside the CE, and explore the effect of
different physical conditions on the possible progenitors of SNe Ia, SN 2006gy-like
transients. Formation of CO WD-NS systems as GW sources
is also studied in this work. In \S 2, we
describe our method to treat the binary physical processes and
parameters in the BPS simulations, such as mass transfer (decided
from critical mass ratio), CE
(different CE efficiency, different binding energy parameter) and
metallicity. Formation and evolution of the CO WDs + intermediate
mass stars including He stars as the progenitors
for SNe Ia are shown in \S 3. We show that two channels in \S 3 could produce one
sub-types of SNe Ia which is the over-luminous / super-Chandrasekhar mass SNe Ia, 2003fg-like SNe Ia (Howell
et al. 2006; Hicken et al. 2007; Scalzo et al. 2010; Hachisu et al. 2012). The possible contribution of mergers
of CO WDs with the He cores of massive stars to the SN 2006gy-like
SLSNe is examined in \S 4. CO WD -NS systems that evolved from the
CO WD + massive star binaries are also investigated in \S 4. The
discussion and summary are in \S 5.

\section{Binary Population Synthesis}
\label{sec:model}

In this work, a large number ($10^7$ ) of binary evolution
calculations have been done by using the updated BSE population synthesis code with Monte Carlo technique
(Hurley et al. 2002;  Kiel \& Hurley 2006; Ablimit et al.
2016; Ablimit \& Maeda 2018). In this updated BPS code, the initial
input parameters of the primordial binaries are set as follows. In the subroutine $sample$ of the code,  the
initial mass function of Kroupa et al. (1993) is adopted for the
primary (the more massive one in a binary) mass,
\begin{equation}
f(M_1) = \left\{ \begin{array}{ll}
0 & \textrm{${M_1/M_\odot} < 0.1$}\\
0.29056{(M_1/M_\odot)}^{-1.3} & \textrm{$0.1\leq {M_1/M_\odot} < 0.5$}\\
0.1557{(M_1/M_\odot)}^{-2.2} & \textrm{$0.5\leq {M_1/M_\odot} < 1.0$}\\
0.1557{(M_1/M_\odot)}^{-2.7} & \textrm{$1.0\leq {M_1/M_\odot} \leq 150$},
\end{array} \right.
\end{equation}
The secondary mass is decided by following the distribution of the initial mass ratio of the secondary to the primary,
\begin{equation}
n(q) = \left\{ \begin{array}{ll}
0 & \textrm{$q>1$}\\
\mu q^{\nu} & \textrm{$0\leq q < 1$},
\end{array} \right.
\end{equation}
where $q=M_2/M_1$, $\mu$ is the normalization factor for the assumed power law distribution
with the index $\nu$. A flat initial mass ratio distribution which are $\nu = 0$ and $n(q)=$constant are utilized in the calculations.
The initial orbital separation, $a_{\rm i}$, is used as
(Davis et al. 2008),
\begin{equation}
n(a_{\rm i}) = \left\{ \begin{array}{ll}
0 & \textrm{$a_{\rm i}/R_\odot < 3$ or $a_{\rm i}/R_\odot > 10^{6}$}\\
0.078636{(a_{\rm i}/R_\odot)}^{-1} & \textrm{$3\leq a_{\rm i}/R_\odot \leq 10^{6}$} \ .
\end{array} \right.
\end{equation}
The uniform initial eccentricity distribution which has a range between 0 and 1 is taken for the calculations.
Initial binary parameters $M_1$ $M_2$ and $a$
within the limits introduced above can be set up with the $\rm n_x$ grid points of parameter $\chi$ logarithmically spaced,
\begin{equation}
\delta {\rm ln\chi} = \frac{1}{\rm n_x - 1}({\rm ln{\chi_{\rm max}}} - {\rm ln{\chi_{\rm min}}}),
\end{equation}
For each set of initial parameters, we evolve the binary system to
an age of the Hubble time, or until it is destroyed. Each phase of the
evolution, such as tidal evolution, angular momentum changes due to mass variations etc.,
is followed in detail according to the algorithms described in Hurley et al.(2002).
Other main revisions in the code for this work are as follows.

For the SN remnant calculation, we adopted the rapid
remnant-mass model of Fryer et al.(2012) in the subroutine \textit{HRDIAG}
of the code, and note that a typo in Fryer et al. (2012) has been
corrected in our code (corrected as $a_1 = 0.25 - \frac{1.275}{(M_1
- M_{\rm proto})}$ in Equation 16). The NS formation by the accretion-induced collapse of a WD
and the possibility of NS formation through electron-capture SN is
taken into account in the work. The wind mass loss
prescription of Vink et al. (2001) is utilized for O and B stars in
different stages (hot stars) in the subroutine \textit{mlwind}
of the updated code. For the wind mass loss of luminous blue variable stars-
$1.5\times10^{-4}\,\dot{M}_\odot\,\rm yr^{-1}$ (Vink \& de Koter
2002) is taken, while default wind mass loss prescriptions of Hurley et al.(2000) are used for other types of stars. The velocity dispersion of
$\sigma_{\rm k} = 265\,{\rm km\,s^{-1}}$ (Hobbs et al. 2005) and $\sigma_{\rm k} = 40\,{\rm km\,s^{-1}}$ (Dessart et al. 2006) are
employed in the Maxwellian distribution of the kick velocity
imparted to the newborn NS formed from CCSNe and
electron-capture SNe, respectively. Note that these prescriptions
would not affect the formation of the CO WD + non-degenerate star systems which never experience SN explosions nor evolve through very massive stars.
However, they can influence the formation of CO WD - NS systems.
A CO or ONeMg WD accretes matter
and grows in mass, and finally WD explodes or collapses to form an NS when the WD's mass reach to
the Chandrasekhar limit mass (1.44 $M_\odot$ is adopted in this work). The accretion efficiency
is important, and the magnetic field also may play key role during the mass transfer process of
the WD binary (e.g., Hillman et al. 2016; Ablimit \& Maeda 2019a, b; Ablimit 2019).
The CO WDs mildly accrete matter and grow in mass when the unstable mass transfer (CE phase) considered.

$Mass~transfer~stability~and~critical~mass~ratio$: The stability of
mass transfer has a decisive role in the binary evolution, and a
binary would evolve into a CE phase if the mass transfer becomes
dynamically unstable. It is not an easy task to simulate the mass
transfer in the BPS or detailed evolution codes (e.g. de Mink et al.
2007; Pavlovskii \& Ivanova 2015). The mass transfer in the BPS code
is usually determined with parameterized and simplified stellar
models. In this study, the critical mass ratio is seen as a key
parameter in the binary evolution which has a crucial role in the
stability of mass transfer. The CE evolution may start if the mass
transfer becomes dynamically unstable when the mass ratio of two
stars is larger than $q_{\rm crit}$, and the CE phase is very
important for the formation of various celestial objects. Hjellming
\& Webbink (1987) showed that $q_{\rm crit}$ varies with the
evolutionary state of the donor star at the onset of RLOF.
We adopt two different prescriptions for $q_{\rm
crit}$ when the donor star is on MS or crossing the Hertzsprung Gap
(HG) in the subroutine \textit{evolv2} (for more details see Ablimit et al. (2016),
Ablimit \& Maeda (2018), and Table 1). In the first model which
is referred to as $q_{\rm crit} = q_{\rm const}$, it is taken as $q_{\rm crit} = q_{\rm const} =
4.0$ (e.g., Hurley et al. 2002). We adopted another prescription of $q_{\rm crit} = q_{\rm cs}$
which is derived by considering the spin of the accretor during the the mass transfer process
(e.g., Ablimit et al. 2016; note that here the mass ratio $q_{\rm cs}$ is the primary to secondary stars).
In the mass transfer phase, even a small amount of mass accretion onto the secondary star from primary star
will make the secondary spin faster and closer to its critical value (Packet 1981). In this model,
the spin of the accretor may reduce the accretion
rate onto the rotating star by a factor of ($1 -\Omega / {{\Omega}_{\rm cr}}$),
where ${\Omega}$ is the angular velocity of the star and ${\Omega}_{\rm cr}$ is its critical
value (e.g., Stancliffe \& Eldridge 2009).
Thus, this rotation-dependent mass transfer makes the mass transfer
accretion rate very low, and most of the transferred matter can be escaped from the binary.
We assume that the material is ejected with the specific
angular momentum of the accretor (e.g., Hurley et al. 2002).
When we consider the non-conservative mass transfer with highly mass loss, the maximal initial
mass ratio of the primary to the secondary stars in primordial
binaries will be $\sim$5-6 (the mass ratio increases with the spin of the accretor), and a
larger number of the primordial binaries will not have the contact phase and can experience stable
mass transfer phases until the primary's envelope is completely
exhausted (see Abdusalam et al. (2020) for similar discussions and results).
If the primordial primary is on the
first giant branch (FGB) or the AGB, we use

\begin{equation}
q_{\rm crit} = 0.362 + \frac{1}{3(1 - {M_{\rm c}}/M)},
\end{equation}
where $M$ and $M_{\rm c}$ are the whole mass and core mass of the
donor star (Hjellming \& Webbink1987). If the mass donors
 are stripped helium-rich MS or helium-rich
giants, $q_{\rm crit} = 3.0$ or 0.784 (e.g., Hurley et al. 2002), respectively.

$Common~envelope~model$: The energy conservation model
(e.g., De Marco et al. 2011; Ivanova
et al. 2013) has been applied for the CE evolution\footnote{There are some other different
mechanisms to change fate of the CE evolution, such as jets (e.g., Shiber et al. 2019)}, and there are
two important but unknown physical parameters in the model which
have been set as constants. They are the CE efficiency ($\alpha_{\rm CE}$) and binding energy
($\lambda$) parameters. Whether the two stars merged to one or
survived from the CE to continue their evolution depend on these two
parameters, and the relation is,
\begin{equation}
\frac{G M_{1,{\rm i}} M_{1,{\rm env}}}{\lambda R_{\rm RL}} =
\alpha_{\rm CE}(\frac{GM_{1,{\rm f}}M_{2,{\rm f}}}{2a_{\rm f}} - \frac{GM_{1,{\rm i}}M_{2,{\rm i}}}{2a_{\rm i}}),
\end{equation}
where $G$ is the gravitational constant;$M_{1,{\rm i}}$, $M_{1,{\rm env}}$ and $M_{1,{\rm f}}$ are the
initial, envelope and final masses of the primary star, respectively; $M_{2,{\rm i}}$, $M_{2,{\rm env}}$
and $M_{2,{\rm f}}$ are the initial, envelope and final masses of the secondary star, respectively;
$R_{\rm RL}$ is the Roche lobe radius of the donor star at the
onset of RLOF.
We first take a constant case $\lambda = 0.5$ for the binding
energy parameter. However, it has been ascertained that the binding energy
parameter changes with the mass and evolutionary stage of the star
(e.g., Dewi \& Tauris 2000). We also use variable $\lambda = \lambda_{\rm w}$ in the subroutine \textit{comenv} of the updated
code by following Wang et al. (2016; see also Klencki et al. (2020)) in which they estimated lambda based
on detailed stellar evolution models. For the CE efficiency, we take $\alpha_{\rm CE} = 1.0$ (e.g., Klencki et al. 2020) in the simulations,
and $\alpha_{\rm CE} = 0.1$ is taken as a pessimistic one for the comparison.

In order to test the effect of the metallicity on the
model outcomes, the initial metallicity of stars is set to be 0.02
and 0.001. Default values of other parameters (Hurley et al. 2002)
are used in the calculation. In total,
we have five models which cover results of other parameter combinations, and see
Table 1 for the main prescriptions taken for the BPS simulations.

$Description~of~the~key~merger~model$: The thermonuclear explosion
of a CO WD following an initial He detonation on the WD surface (the double detonation scenario for SNe Ia) has
been discussed as a popular scenario for SNe Ia (e.g. Livne 1990;
Iben \& Tutukov 1991; Woosley \& Weaver 1994; Fink et al. 2010;
Townsley et al. 2012; Jiang et al. 2017; De et al. 2019). A CO
WD merger with another CO WD with the He-rich shell or merger
between a CO WD and a He WD has been suggested for triggering the SN
Ia explosion through the He detonation (Shen et al. 2018; Tanikawa
et al. 2019). In this work, we assume that a merger of a CO WD with
a He core of the companion star during the CE phase could initiate
the SN explosion, which we term as the core merger
detonation (CMD) model. In this hypothesis, a thick He
layer would be accumulated on the surface of the CO WD following the
rapid accretion during the spiral-in process between the CO WD and He
core of the H-rich non-degenerate star after the CE event,
then conditions for the initial He detonation and stronger
second detonation inside the WD could be achieved. Due to the short
timescale and lack of direct observational evidences for the CE
phase, what happens inside the CE and merger outcomes from the CE \textbf{are}
poorly known. Besides, the exact conditions (mass of the He layer,
temperature, density of the CO WD, etc.) for the He detonation for
the SN explosion are also not well established. Thus, we simply
assume that at least a $\geq 0.6M_\odot$ core of a star (e.g.
Bloecker 1995; Soker 2015) and a CO WD with a critical mass $\geq 0.9
M_\odot$ or $\geq 1.0 M_\odot$ inside the CE would be needed to
trigger the SN (e.g., Shigeyama et al. 1992; Sim et al. 2010; Ruiter et al. 2011).
We set a reasonable critical mass for the CO WD for the He detonation-
triggered SN explosion, because a massive enough WD is helpful for the
accreting He materials to experience the stronger shock heating,
and a massive enough WD is dense enough to have stronger detonation.

The CO WD + He-rich non-degenerate star systems are descendants of CO WD + H-rich normal star binaries,
and they also may evolve into the CE phase. In the CO WD $+$ He star binary, the core of He star may not
or may further evolve (it depends on evolution conditions/stellar evolution stage when the CE occurs), thus
the CO WD may have rapid accretion of helium or carbon during the merger process \textbf{in a} CE phase.
We simply assume that the rapid accretion of a helium-rich or carbon-rich material or
He $+$ C material as in the hybrid double degenerate model (Perets et al. 2019) onto a CO WD in the merger process may trigger SNe Ia.
Regarding the uncertain issues in the explosion, evolution mechanisms, and limitations
in the one-dimensional (1D) BPS simulation, we just consider that the thermonuclear SN may happen in the
CO WD $+$ He star binary during the CE phase if the CO WD mass satisfies the critical value showed above.

$Uncertainties$: We developed the BPS code with some new physical ingredients as introduced above,
which can be helpful to our understanding of the binary evolution as origins of peculiar transients. However,
there are still some common issues in this work comparing to similar previous works,
such as not-well constrained parameters/processes (i.e., $\alpha_{\rm CE}$) in the BPS and limitations on
the explosion mechanisms of SNe (i.e., exact mass of the CO WD to explode).
The commonly accepted and simplified/paramitized ways are adopted in the BPS,
which may different from more detailed modeling (e.g., MESA modeling), and the predicted rate from the
systematic study with large populations is always one of the important/meaningful outcomes of such studies.
Another important question is whether the merger process is a "fast merger" or a "slow merger".
The idea of the "fast merger" is that the spiral-in occurs on
a dynamical timescale and the companion is dynamically disrupted and may lead to rapid accretion onto the compact core and trigger the detonation.
In the "slow merger" case, the He core may fills its RL and starts mass transfer onto the compact
core during the spiral-in process, and this would have a longer delay time to trigger the detonation. It still remains unclear that what would happen inside the CE.
More observational constraints and theoretical works need to be placed for
binary evolutionary scenarios and explosion mechanisms for SNe in the future.

\section{Mergers of CO WDs with Cores of Intermediate Mass Stars and He Stars within the CE Phase: Possible Explosions of SNe Ia}

There are 467918 CE events evolved from $1\times10^7$ initial MS-MS
binaries in our standard BPS simulation (model 1), and this CE event number is
consistent with results of similar works (e.g., Howitt et al. 2020).
Among these CE events, CO WD CE events with H-rich normal stars and
He stars are investigated in this work. From interacting MS-MS
binaries, there are several main evolutionary pathways to produce a
primary CO WD with a non-degenerate companion system. We briefly
introduce the one main formation channel of CO WD + evolved intermediate-mass
stars ($M_2 < 8.0M_\odot$): The primary star with relatively higher mass
in a MS-MS binary evolves first, loses its mass via RLOF to the
secondary star and becomes a CO WD. The secondary starts its
evolution after the CO WD formed, and the unstable RLOF mass
transfer may occur when it becomes a HG star, a red giant (RG)
star or an AGB star\footnote{Note that the WD binary
evolution with the dynamically stable mass transfer and WD+MS
binaries (see Willems \& Kolb (2004) for more details) are out of scope
of this work.}. The envelope of the secondary engulfs the CO WD due
to the unstable and rapid RLOF mass transfer, then a system composed
of the CO WD + the core of the secondary inside the CE is formed.
Figure 1 shows the mass distributions of all CO WDs + H-rich stars
and CO WDs + He stars just before entering the CE evolutions, and
also shows initial masses which cause those CE events under
our standard model. The CO WD + H-rich star CE event channel is
realized for the following typical ranges in the initial binary
parameters; the initial primary mass $M_{1,\rm
i}\sim1.0-10.0M_\odot$, the initial mass ratio $q=M_{2,\rm
i}/M_{1,\rm i}\sim0.15-1.0$; and the initial orbital period $P_{\rm
orb, i}\sim0.01-93.6$ days. The mass range of the WD in this channel
is between $\sim 0.3$ and $\sim 1.3 M_\odot$ while it is between
$\sim 0.5$ and $\sim 8.0 M_\odot$ for the companion star (see Figure
1).

Event rates and delay time ranges of the CE systems which contain CO
WDs + non-degenerate H-rich stars ($M_2 < 8.0M_\odot$) under different
conditions and five different models are summarized in Table 2. The
rates of the all CE events (CO WDs + non-degenerate H-rich stars
($M_2 < 8.0M_\odot$)) clearly change with the $\alpha$ parameter,
metallicity and mass transfer model. The pessimistic value of $\alpha=0.1$ in model 2
causes more mergers during the CE events of non-degenerate stars and
non-degenerate stars, therefore it reduces the number of CO WD
binaries with CE phases from 0.44\% to 0.27\%. The central
temperatures and densities are higher, and more hydrogen burns
stably for the low-metallicity MS stars, and it leads to larger core
masses and different lifetimes during the evolution. Thus, more CO
WD binaries can be formed with the low metallicity condition. The rotation-dependent (highly nonconservative)
mass transfer model adopted in this work during the primordial binary evolution
allows a large amount of the primordial binaries to
experience stable mass transfer, thus this reduces the number of CO WD binaries with CE events. About
0.08\% - 0.12\% CO WDs merge with cores of H-rich stars at different
evolutionary stages inside the CE, and mergers (numbers include
delay times) at different evolutionary stages such as HG, RG or AGB
\footnote{The initial conditions, such as initial masses and initial separations,
are important in determining at which evolutionary phase of the companion the WD enters its
envelope (e.g., Hurley et al. 2002). These conditions are extensively studied in the literature and we do not give the details here.}
are affected by the mass transfer (critical mass ratio), $\alpha$
and $\lambda$ parameters as these physical parameters are related
to stellar structure and evolution. It is worth to note that at
least 20\% - 40\% mergers of CO WDs and cores of intermediate
mass stars have combined core masses larger than 2$M_\odot$ (Figure 2).
Thus, they might be the progenitors of over-luminous SNe Ia.

We investigate the possible CMD explosions
triggered by mergers between CO WDs and He cores of non-degenerate
intermediate stars during CE events by considering the critical
masses introduced above.
 They would be the origins for SNe Ia-CSM or/and overluminous SNe Ia,
and their rates are given in Table 2. This channel includes products
of the CD model when the WD CE core merges with an AGB star.  Figure
2 displays property distributions of CO WD binaries which lead to
the CMD event. The mass transfer model (model 5) clearly affects all CE event numbers,
mass and orbital period distributions due to the mass accretion, mass loss
and corresponding angular momentum evolution. In the mass transfer simulation of model 5,
the larger mass loss (corresponding angular momentum loss) of the
primordial binary due to the mass ratio (see section 2) makes the mass transfer stable,
thus more primordial binaries can avoid CE phases (and have different orbital periods)
comparing to the other model. The $\alpha$ parameter
(model 2) causes different event numbers and distributions including delay times
(Table 2), because the less energy budget contributed for expelling the CE
in the model 2 with $\alpha=0.1$ comparing to model 1 with $\alpha=1.0$.
Heavier cores are formed
with the low metallicity (model 3) due to the same reason discussed above. Two different prescriptions of
$\lambda$ (model 1 and model 4) give similar mass distributions (Figure 2) while
CE event numbers are different. $\lambda$ is not the physical parameter for the single stellar
evolution, so it does not affect the stellar features such as mass of the star.
It is a parameter in the CE evolution
of the binary. The more energy (i.e. internal energy) that is available in the $\lambda$
prescription of model 4 for the CE ejection (see above section for more details) implies more
successful CE ejections in the primordial binaries. This in turn reduces the number of CO WD CE systems in later evolution phases.

Comparing with the observed/time-integrated rate of SNe Ia which is about $10^{-3}{M_\odot}^{-1}$
(e.g. Maoz et al. 2014; Maoz \& Graur 2017; Frohmaier et al. 2018), all predicted rates from this channel (Table 2)
are lower than the observational rate. However, the rate of model 4 is consistent
with the predicted SNe Ia-CSM rate by Soker (2019) and the fraction of known SNe Ia-CSM (roughly about few$\times 10^{-5}{M_\odot}^{-1}$-$10^{-4}{M_\odot}^{-1}$)
among all observed SNe Ia available to date (e.g., Li et al. 2011; Silverman et al. 2013), and it implies that
the treatment of the binding energy in this work is physically motivated one. Besides, there are a few robust
candidates for the He-detonation triggered SNe Ia reported so far, and the observationally inferred rate of
such events which is estimated as 0.5\% of the SN Ia rate (Jiang et al. 2017; De et al. 2019) can be reproduced by this work.
Delay times under conditions 1 and 2 (see the caption of Table 2 for two conditions) for mergers tend to have the shorter delay times than $\sim$ 2 Gyr in our results (see Table 2) due to the evolution timescales,
and they can be longer if the "slow merger" inside the CE is considered.

The H-rich envelope can be stripped away if a binary system survives
from the CE phase, and a He star with a CO WD can be formed. If the
He star begins unstable RLOF mass transfer, the core of the He star
and CO WD enter the CE. The CO WD + He star CE event channel is
realized for the following ranges in the initial binary parameters:
the initial primary mass $M_{1,\rm i}\sim1.2-7.2M_\odot$, the
initial mass ratio $q=M_{2,\rm i}/M_{1,\rm i}\sim0.25-1.0$ and the
initial orbital period $P_{\rm orb, i}\sim0.02-47.1$ in days. The
mass range of the WD in this channel is between $\sim 0.5$ and $\sim
1.35 M_\odot$ while the He stars' mass is distributed between $\sim
0.4$ and $\sim 3.4 M_\odot$ (see Figure 1). In this channel, we
consider mergers between cores of He stars and CO WDs with masses
0.9 or 1.0 $M_\odot$ with the CE event that produce SNe Ia. From the
rates ($0.0 - 1.61\times10^{-5}{M_\odot}^{-1}$) shown in Table 2, we
can see that only at most $\sim$10\% of all CE events of CO WDs $+$
He stars end up with mergers under the condition $M_{\rm WD} \geq
0.9M_\odot$. In Figure 3 we demonstrate mass relations for CO WDs,
He stars and their cores for model 1 which has the highes rate, and
results show that mass distributions have relatively narrow ranges.
These results suggest that the CO WD that merges with the core of
the He star companion may not play an important role as much as
mergers with other different companions. However, this might be the
origin for some rare transients (peculiar SNe Ia like over-luminous
SNe Ia or SNe Iax (see Foley et al. (2013) for the subclass SNe Iax)) which need further
investigations. Our result suggests that the He accretion and
thermonuclear explosion, the so-called double detonation scenario in
SD systems, from CO WDs + He stars that survive from the CE can
contribute to SNe Ia more than those in the merger case.

\section{CO WD and Massive Star Binaries with the CE Evolution}

In our standard model, there are 1232 CO WD + massive star ($>8M_\odot$) systems
which experience the CE phase.
The primordial MS/MS binaries of these systems have the following
typical initial parameters: the initial primary mass $M_{1,\rm
i}\sim5.3-10.0M_\odot$, the initial secondary mass $M_{2,\rm
i}=\sim2.6-7.3M_\odot$ (Figure 4) and the initial orbital period in
a range of $P_{\rm orb, i}\sim0.1-40.5$ days. The mass relation of
1232 CO WDs and companion stars at the onset of the CE are given in
Figure 4.

\subsection{Mergers of CO WDs with Cores of Massive Stars inside the heavy CE: Possible Progenitors for SLSNe like SN 2006gy}

In the standard model (model 1), 505 and 209 CO WDs merge with the He cores of
the massive star companions inside the CE under the condition of
the critical CO WD mass of 0.9 and
1.0$M_\odot$ (Table 3), respectively. The mass relations between CO
WDs and massive stars, and relations between CO WDs + cores of
companions and massive stars' envelopes for the mergers with $M_{\rm
WD} \geq 0.9M_\odot$ are also shown in Figure 4. It can be seen from the figure that
a system can indeed have the envelope as massive as $> 6.0M_\odot$. If the CO WD merges
with the non-degenerate He core of the massive star inside the CE, the accreted helium-rich material
onto the CO WD ignited and started the detonation, then thermonuclear explosion is triggered following the core merger,
thus the detonation process is different from the CD model which is suggested as mergers of two degenerate cores for the SNe Ia.
Two main evolutionary pathways are demonstrated in Figures 5 and 6.
Initially, most secondary stars in these evolutionary pathways are
not massive stars. They become O/B type massive stars due to mass
accretion from the primary stars, and the angular momentum
transported with the accreted mass may spin the secondary up to be
fast rotating Be stars (Waters et al. 1988; Pols et al. 1991;
Willems \& Kolb 2004; Ablimit \& L$\ddot{\rm u}$ 2013).
The increasing number of candidates by the
recent X-ray binary observations (e.g., SWIFT and eROSITA) (Coe et
al. 2020; Haberl et al. 2020a,b) show that this kind of WD binaries
may not be so rare as suggested by previous works (e.g., Li et al.
2012; Cracco et al. 2018). After the CO WDs are formed by going
through two RLOF mass transfers, the massive star companions evolve
and fill their RL at the evolution stages of the HG and core helium
burning (CHeB), and the mass transfers will be unstable
as the mass ratios of CO WDs/massive companions are low. Thus, the
massive giant branch stars' envelopes engulf CO WDs and He cores,
and a merger would happen inside the CE if the CE ejection failed.
The delay times are relatively short which is between 34 and
$\sim$120 Myr. The mergers with HG stars are much more common than
mergers with CHeB stage stars. With a standard BPS simulation, we
have shown that the created CSM (Figure 4) and the rate (Table 3) of CMD SN explosions in
CO WD + massive star systems can be comparable with the
observational constraints of SN 2006gy (Jerkstrand et al. 2020; see below for the rate comparison).

The rates and delay times of CO WDs + massive star core mergers
under different models are given in Table 3. Figure 7 shows the
property distributions of CO WD + massive star CE mergers with
$M_{\rm WD} \geq 0.9M_\odot$ under different models. Comparing the
effect of physical parameters on properties of CO WD + intermediate
star CE mergers, the metallicity with the massive stars case has a
very clear influence on the mass distribution of cores due to the
fact that a heavier core can be formed with low metallicity (see above sections for the reason).
Of course, the mass transfer and $\lambda$ also play an important role in the formation
of CO WD/ Massive star binaries with the CE phase due to the reasons as discussed before. In any case, the combined
masses of CO WDs and He cores of massive stars and H-rich envelope
masses are always larger than 2.5$M_\odot$ and 6.0$M_\odot$,
respectively. Derived envelope masses, core masses and rates
imply that the CO WD + massive star core
mergers could be progenitors of SN 2006gy-like type II SLSNe. Observationally,
there is possibility of mixing the SNe Ia and type II SLSNe, and we
assumed that mergers between CO WDs and He core of massive stars can trigger type II SLSNe.
Based on Quimby et al. (2013), we roughly take the rate of type II SLSNe as $\sim 10^{-5}{\rm yr}^{-1}$,
but note that the rate of type II SLSNe is not observationally well-determined yet.
If we adopt the star formation rate of 2 $M_\odot \, {\rm yr}^{-1}$ and binary fraction of 0.7,
predicted rates ($\sim 1.19-7.7\times 10^{-5}{\rm yr}^{-1}$) of models (mode 1, 4 and 5) are comparable with the observation.

\subsection{Survivors from the heavy CE: Formation of CO WD-NS Systems as the Gravitational Wave Sources}

If the energy budget of the system is enough to expel the CE in the
CO WD + massive star CE systems introduced in the above section, the
CO WD + non-degenerate star binary will be formed in a significantly
reduced orbit, and no clear accretion occurs during the CE (Ivanova
et al. 2013). Given the reduced orbit, the secondary fills its RL
and starts the mass transfer, and the accretion onto the WD depends
on the retention efficiency and mass transfer rate (e.g. Bours et
al. 2013). After a series of evolutions, the secondary may undergo a
SN explosion and collapse to an NS. The eccentricities of the
post-SN orbits span the full range of (0-1) with natal kicks in our
simulations (see Section 2). Formed CO WD - NS systems can be merge
in a Hubble time due to the very small orbital period reduced by the
CE ejection. Table 3 shows the formation rates of CO WD - NS systems
through the channel introduced above. Excluding the model with a
pessimistic value of $\alpha$ (all cores merge inside the CE), the
rate changes from 0.54 to 2.83$\times10^{-5}{M_\odot}^{-1}$ under
different models. Note that here we focused on the reverse formation
pathway that evolved from CO WD + massive star CE systems, while
there are several pathways to form NS-WD systems (e.g., Sabach \&
Soker 2014). More CO WD - NS systems can be formed in our model 4 and 5,
because more primordial binaries can avoid the CE evolution due to the highly
non-conservative mass transfer model (model 5), and the $\lambda$ prescription
used in model 4/5 contributes more energy to expel the CE (more systems survive with reduced close orbits).

The merger of NS-WD systems would have various observable outcomes,
and the GW signal from it is one of the good sources for the next
generation GW detectors. In a CO WD- NS system, the orbital motion
and angular momentum evolution with the GW emission (according to
general relativity) will lead to merger of the two objects at the
final stage of the inspiraling process. The merger timescale (Peters
\& Mathews 1963; Peters 1964; Lorimer 2008; Kuerban et al. 2020) is,

\begin{equation}
t = 9.88\times10^{6}(\frac{P_{\rm orb}}{ 1 \,\rm hr})^{8/3} (\frac{\mu}{1\,M_\odot})^{-1} (\frac{M}{1\,M_\odot})^{-2/3}\, {\rm yr},
\end{equation}
where $\mu = M_1M_2/(M_1 + M_2)$ and $M = M_1 + M_2$ are the reduced
mass and total mass of the system ($M_1$ and $M_2$ are the masses of
two objects in a binary system), respectively. The GW strain will
increase with time during the merging process. If the distance of
the binary system with respect to us is $d$, then the strain
amplitude of the GW (Peters \& Mathews 1963) is,

\begin{equation}
h = 5.1\times10^{-23}(\frac{P_{\rm orb}}{ 1 \,\rm hr})^{-2/3} (\frac{M_{\rm ch}}{1\,M_\odot})^{5/3} (\frac{d}{10\,{\rm kpc}})^{-1},
\end{equation}
where $M_{\rm ch} = (M_1M_2)^{3/5}/(M_1 + M_2)^{1/5}$ is the chirp
mass. The GW frequency $f$ is twice the orbital motion frequency.
The derived GW strain and frequency from merger signals of CO WD -
NS systems formed in models 1 (lowest rate) and 5 (highest rate) are
shown in Figure 8. About 30\% and 75\% of the whole CO WD - NS
populations from models 1 and 5 are detectable by the upcoming GW
detector LISA.

\section{Discussion and Summary}

We explored the formation and end products of binary CO WDs with
non-degenerate companion stars which evolve into the CE phase by
considering a number of physical elements in this BPS simulation. In
particular, we investigated mergers between CO WDs and cores of
non-degenerate stars inside the CE as possible progenitors of
peculiar SNe with the proposed CMD scenario. We examined rates, delay times, properties of
progenitors of these events, and their dependence on mass transfer, metallicity and
CE evolution. Different BPS
models coupled with Monte Carlo method give
relative statistical errors (Ablimit et al. 2016; Ablimit \& Maeda 2018).
The difference in results of this work is mainly caused by the CE and mass transfer models.
Future observational results may provide more direct evidences to constrain the CE evolution and
mass transfer in the binary evolution. Predicted rates from the CMD model in this work
(especially results of our model 4) are basically consistent with observational results of special transients.

For the CO WD + intermediate mass star channel,  $\sim$0.27 \% -
0.44\% of all $10^7$ systems in a simulation have evolved toward CO WD +
intermediate mass star binaries with the CE phases under different
models. Among all CO WD + intermediate mass star CE events, $\sim$14\% - 25\% of CO WDs merge with cores of companion
stars. The model with the $\alpha=0.1$ fails to survive from the CE and only has mergers, because very low energy contribute to expel the CE in this case.
The merger under two conditions introduced in the caption of Table 2 may trigger SNe Ia, and
corresponding rates are (0.22 - 8.77)$\times10^{-5}{M_\odot}^{-1}$ and
(0.0 - 4.7)$\times10^{-5}{M_\odot}^{-1}$ for lower and higher WD mass
cases (Table 2), respectively. This CMD scenario here includes
the CD model as an outcome of the core merger between a CO WD and an
AGB star, and roughly 24\%-55\% of our rates are the rates derived from the CD scenario. Delay times from the star formation to the merger are mainly
distributed in a range of $\sim$0.1-2.5 Gyr. Our results show that
the merger systems with $M_{\rm WD}\geq0.9M_\odot$ could produce the
CSM with mass larger than 2$M_\odot$, and this could account for a
certain fraction of observed SNe Ia interacting with CSM. 0.002\% - 0.024\% of all $10^7$ binaries are CO WD + He star CE events, and core merger
rates from this channels are too low to
explain the normal SNe Ia. However, the merger between CO WDs and
core of He stars may initiate peculiar SN transients (variant of SNe
Ia) such as SNe Iax and overluminous SNe Ia. Heavier core masses ($>2M_\odot$) of mergers imply that the two evolution pathways, that of CO WD + intermediate mass star CE events and that of CO WD + He star CE events, may be origins for overluminous (super-chandrasekhar) SNe Ia, and predicted rates for this peculiar SNe Ia
have a range of $\sim$(0.02 - 3.31)$\times10^{-5}{M_\odot}^{-1}$ according to different models.

Another important aim of the work is to investigate the possible
origins for SN 2006gy-like type II SLSNe with different binary evolution models.
Observational results of Jerkstrand et al. (2020) imply that
a merger of a CO WD with a He core of a massive star (Sabach \& Soker 2014) may be the
progenitor of SN 2006gy. The rapid He accretion onto the CO WD
during the spiral-in merger process inside the CE may trigger SN-like
transients which is termed as the CMD
scenario in this work. By performing BPS simulations with different
physical models, we explored the binary evolutions toward CO WD +
massive star CE events. In general, only $\sim$0.012\%
primordial binaries formed CO WDs + massive star binaries with the
CE, and rates of core mergers among them are different according to
adopted physical parameters in the binary evolution as summarized in
Table 3. They have relatively short delay times. The derived rates
for SN 2006gy-like type II SLSNe in this channel are at least two
orders of magnitude lower than the rate of observed CCSNe.
The helium-burning core of a supergiant and the ejected envelope can be a factor
of ten larger than that required for a lower mass AGB star (see results in above sections), therefore
CE simulations involving massive donor stars offer challenges beyond those
required for low- and intermediate-mass systems.

Survivors from CO WD + massive star CE events eventually can become
CO WD - NS systems in very close orbits. The rates of CO WD - NS
binaries are (0.54 - 2.83)$\times10^{-5}{M_\odot}^{-1}$ under different
models except model 2. About 30\% - 75\% of CO WDs - NS systems that
formed through the CO WD + massive star CE channel would emit GW
signatures during the spiral merger process, which can be detected
by LISA.

Our results once again show that the stability and nature of the
mass transfer, CE evolution and metallicity are very important in
binary stellar evolution. Whether a binary experiences a CE
evolution or not, whether the cores merge or survive from the CE,
and how stellar evolution (e.g. core mass) depends on metallicity
are described in this work. This is a clear demonstration that pre-CE
binary interactions can play an important role in the outcome of the CE.
Rates and properties are affected by
these physical ingredients, especially with extreme conditions.
 This
work can be used for future observations to improve our
understanding of binary stellar evolution and the nature of
transient events. In our future work, we will study ONeMg WD binaries
with CE evolution as possible origins for peculiar transients and objects (Ablimit et al. 2021).

\section*{Acknowledgements}

I thank the reviewer of the paper for the careful reading and useful comments to improve the manuscript.
I also thank James Wicker and Takashi Moriya for the support and discussions. This work was funded by the LAMOST FELLOWSHIP which
is supported by Special Funding for Advanced Users, budgeted and
administered by the Center for Astronomical Mega-Science, CAS.

\textbf{Data Availability:} The data underlying this article
will be shared on reasonable request to the corresponding author.






Abdusalam, K., Ablimit, I., Hashim, P., L$\ddot{\rm u}$, G.-L. et al. 2020, ApJ, 902, 125

Ablimit, I. \&  L$\ddot{\rm u}$, GuoLiang, 2013, SCPMA, 56, 663

Ablimit, I., Maeda, K. \& Li, X.-D., 2016, ApJ, 826, 53

Ablimit, I., \& Maeda, K., 2018, ApJ, 866, 151

Ablimit, I., \& Maeda, K. 2019a, ApJ, 871, 31

Ablimit, I., \& Maeda, K. 2019b, ApJ, 885, 99

Ablimit, I. et al. 2021, in preparation

Bloecker, T., 1995, A\&A, 299, 755

Bours, M. C. P., Toonen, S., \& Nelemans, G. 2013, A\&A, 552, A24

Briggs, G. P., Ferrario, L., Tout, C. A. et al., 2015, MNRAS, 447, 1713

Canals, P., Torres, S., \& Soker, N. 2018, MNRAS,
480, 4519,

Coe, M. J., Kennea, J. A., Evans, P. A., \&
Udalski, A. 2020, MNRAS, 497, L50

Cracco, V., Orio, M., Ciroi, S., et al. 2018, ApJ,
862, 167

De, K., Kasliwal, M. M., Polin, A., et al. 2019,
ApJL, 873, L18

De Marco, O., Passy, J.-C., Moe, M. et al. 2011, MNRAS, 411, 2277

Eldridge, J. J., Izzard, R. G. \& Tout, C. A. 2008, MNRAS, 384, 1109

Frohmaier, C., Sullivan, M., Maguire, K., \& Nugent, P. 2018, ApJ, 858, 50

Fink, M., Ropke, F. K., Hillebrandt, W., Seitenzahl,
I. R., Sim, S. A., Kromer, M., 2010, A\&A, 514, A53

Foley, R. J., Challis, P. J., Chornock, R., et al. 2013, ApJ, 767, 57

Fryer, C. L., Belczynski, K., Wiktorowicz, G., et al. 2012, ApJ, 749, 91

Gal-Yam, A. 2019, AR\&A, 57, 305

Haberl, F., Maitra, C., Greiner, J., et al. 2020a,
The Astronomer's Telegram, 13709, 1

Haberl, F., Maitra, C., Greiner, J., et al. 2020b, The Astronomer's Telegram, 13789, 1

Hachisu, I., Kato, M. \& Nomoto, K. 2008, ApJ, 683, 127

Hachisu, I., Kato, M., Saio, H. \& Nomoto, K. 2012, ApJ, 744, 69

Hamers, A. S., Pols, O. R., Claeys, J. S. W. \& Nelemans, G. 2013, MNRAS, 430, 2262

Hamuy, M., Phillips, M. M., Suntzeff, N. B., et al. 2003, Natur, 424, 651

Han, Z. \& Podsiadlowski, Ph. 2004, MNRAS, 350, 130

Howitt, G., Stevenson, S., Vigna-Gomez, A. r., et al. 2020, MNRAS, 492, 3229

Hicken, M., Garnavich, P. M., Prieto, J. L., et al. 2007, ApJL, 669, L17

Hjellming, M. S.,\& Webbink, R. F. 1987, ApJ, 318, 794

Hobbs, G., Lorimer, D. R., Lyne, A. G., \& Kramer, M. 2005, MNRAS, 360, 974

Howell, D. A., Sullivan, M., Nugent, P. E., et al. 2006, Natur, 443, 308

Hurley, J. R., Tout, C. A. \& Pols, O. R. 2002, MNRAS, 329, 89

Iben, I., Jr, Tutukov, A. V., 1984, ApJS, 54, 335

Iben, I., Jr, Tutukov, A. V., 1991, ApJ, 370, 61

Inserra, C., Smartt, S., Jerkstrand, A., et al. 2013, ApJ, 770, 128

Ivanova, N., Justham, S., Chen, X., et al. 2013, A\&A Rev., 21, 59

Jerkstrand, A., Maeda, K. \& Kawabata, K. S. 2020, Science, 367, 415

Jiang, J.-A., Doi, M., Maeda, K., et al. 2017,
Nature, 550, 80

Kashi, A. \& Soker, N. 2011, MNRAS 417,1466

Klencki, J., Nelemans, G., Istrate, A.-G. \& Chruslinska, M., 2020, arXiv:2006.11286

Kiel, P. D. \& Hurley, J. R. 2006, MNRAS, 369, 1152

Kuerban, A., Geng, J.-J., Huang, Y.-F. et al. 2020, ApJ, 890, 41

Langer, N., Deutschmann, A., Wellstein, S. et al. 2000, A\&A, 362, 1046

Li, X.-D. \& van den Heuvel, E. P. J. 1997, A\&A, 322, L9

Li, K. L., Kong, A. K. H., Charles, P. A., et al.
2012, ApJ, 761, 99

Livne, E., 1990, ApJ, 354, L53

Livio, M. \& Riess, A. G. 2003, ApJ, 594, L93

Livio, M. \& Mazzali, P., 2018, Phys. Rep., 736, 1

Li, W., Leaman, J., Chornock, R., et al. 2011, MNRAS, 412, 1441

L$\ddot{\rm u}$, G., Zhu, C., Wang, Z., \& Wang, N. 2009, MNRAS, 396, 1086

Maeda, K. \& Terada, Y. 2016, IJMPD, 253002

Maoz, D., \& Graur, O. 2017, ApJ, 848, 25

Maoz, D., Mannucci, F., \& Nelemans, G. 2014, ARA\&A, 52, 107

Moriya, T. J., Blinnikov, S. I., Tominaga, N.,
et al. 2013, MNRAS, 428, 1020

Nomoto, K., 1982, ApJ, 257, 780

Ofek, E. O., Cameron, P. B., Kasliwal, M. M., et al. 2007, ApJL, 659, L13

O'Shaughnessy, R. \& Kim, C. 2010, ApJ, 715,
230,

Packet, W. 1981, A\&A, 102, 17

Paczy$\acute{\rm n}$ski, B. 1976, in Structure and Evolution of Close Binaries, ed. P. P.
Eggleton, S. Mitton, \& J. Whelan (Dordrecht: Kluwer), 75


Pastorello, A., Smartt, S., Botticella, M., et al. 2010, ApJL, 724, L16

Pavlovskii, K. \& Ivanova, N. 2015, MNRAS, 449, 4415

Perets, H. B., Zenati, Y., Toonen, S., Bobrick, A. 2019, arXiv:1910.07532

Pols, O. R., Cote, J., Waters, L. B. F. M. \& Heise, J. 1991, A\&A, 241, 419

Quimby, R. M., Kulkarni, S. R., Kasliwal, M. M. et al. 2011, Nature, 474, 487

Quimby, R. M., Yuan, F., Akerlof, C., \& Wheeler,
J. C. 2013, MNRAS, 431, 912

Ruiter, A. J., Belczynski, K., Sim, S. A. et al., et al. 2011, MNRAS, 417 408

Sabach, E. \& Soker, N. 2014, MNRAS 439, 954

Scalzo, R. A., Aldering, G., Antilogus, P., et al. 2010, ApJ, 713, 1073

Shiber, S., Iaconi, R., De Marco, O. \& Soker, N. 2019, MNRAS, 488, 5615

Shigeyama, T., Nomoto, K., Yamaoka, H., Thielemann F.-K., 1992, ApJ, 386,
L13

Silverman, J. M., Nugent, P. E., Gal-Yam, A.,
et al. 2013, ApJS, 207, 3

Sim, S. A., Ropke, F. K., Hillebrandt, W. et al. 2010, ApJ, 714, L52

Smith, N., Li, W., Foley, R. J., et al. 2007, ApJ, 666, 1116

Soker, N. 2011, eprint arXiv:1109.4652

Soker, N., Kashi, A., Garcia-Berro, E. et al. 2013, MNRAS, 431, 1541

Soker, N. 2015, MNRAS, 450, 1333

Soker, N. 2019, eprint arXiv:1912.01550

Soker, N. 2019, MNRAS, 490, 2430

Sparks, W. M. \& Stecher, T. P. 1974, ApJ, 188, 149

Stancliffe, R., \& Eldridge, J. 2009, MNRAS, 396, 1699

Tauris, T. M. \& Sennels, T. 2000, A\&A, 355, 236

Toonen, S., Nelemans, G. \& Portegies Zwart, S. 2012, A\&A, 546, 70

Townsley, D. M., Moore, K., Bildsten, L., 2012, ApJ, 755, 4

Tutukov, A.V. \& Yungelson, L.R. 1979, Acta Astron., 29, 665

Tutukov, A.V. \& Yungelson, L.R. 1993, Astronomy Reports, 37, 411

van Paradijs, J., Kouveliotou, C., Wijers, R.A.M.J. 2000,
ARA\&A, 38, 379

Vink, J. S., de Koter, A., \& Lamers, H. J. G. L. M. 2001, A\&A, 369, 574

Vink, J. S., \& de Koter, A. 2002, A\&A, 393, 543

Wang, B., Zhou, W. H., Zuo, Z. Y., et al. 2017,
MNRAS, 464, 3965,

Wang, C., Jia, K., \& Li, X.-D. 2016, RAA, 16h, 9

Waters, L. B. F. M., Taylor, A. R., van den Huevel, E. P. J., Habets, G. M. H. J. \& Persi, P. 1988, A\&A, 198, 200

Webbink, R. F. 1984, ApJ, 277, 355

Willems, B. \& Kolb, U. 2004, A\&A, 419, 1057

Woosley, S. E., Weaver, T. A., 1994, ApJ, 423, 371

Woosley, S. E., Blinnikov, S., \& Heger, A. 2007,
Nature, 450, 390,

Yungelson, L. R., Livio, M., Tutukov, A. V., \& Saffer, R. A. et al. 1994, ApJ, 420, 336

Zapartas, E., de Mink, S. E., Izzard, R. G., et al. 2017, A\&A, 601, A29


\begin{table}

\caption{Different models in our calculation. $\alpha_{\rm CE}$, $\lambda$, $q_{\rm crit}$ and $z$ are the CE efficiency parameter, binding energy parameter in the CE evolution ($\lambda_{\rm w}$ follows Wang et al.(2016)), critical mass ratio and metallicity, respectively.}

\begin{tabular}{c c c c c c}
 \hline\hline
Model & $\alpha_{\rm CE}$ & $\lambda$ & $q_{\rm crit}$ & $z$ \\
\hline
 mod1 & 1.0 & 0.5 & $q_{\rm const}$ & 0.02 \\
 mod2 & 0.1 & 0.5 & $q_{\rm const}$  & 0.02 \\
 mod3 & 1.0 & 0.5 & $q_{\rm const}$ & 0.001 \\
 mod4 & 1.0 & $\lambda_{\rm w}$ & $q_{\rm const}$ & 0.02 \\
 mod5 & 1.0  & $\lambda_{\rm w}$ & $q_{\rm cs}$ & 0.02 \\

\hline
\end{tabular}
\end{table}

\clearpage

\begin{table}

\begin{center}
\caption{Calculated rates ($R$) of CO WD/non-degenerate star CE events under five different models (in $10^{-5}$ ${M_\odot}^{-1}$)}

\begin{tabular}{lccccccc}
 \hline\hline
CE events of CO WDs/companion stars & $R1$ & $R2$& $R3$ & $R4$ & $R5$ \\

\hline
a:CO WDs/normal stars (all CE) &441.30 & 271.18 & 543.46  & 405.82 & 363.58 \\
b:CO WDs/HG stars (merger) & 37.03 & 18.87 & 48.56  & 34.38 & 0.00 \\
c:CO WDs/RG stars (merger) & 51.49 & 75.23 & 26.30   & 86.83 & 89.85 \\
d:CO WDs/AGB stars (merger) & 0.33 & 11.17 & 2.84    & 0.00 & 0.00  \\
e:CO WDs/normal stars (merger con1) & 5.37 & 8.77 & 7.86  & 8.75 & 0.22\\
f:CO WDs/normal stars (merger con2) & 0.22 & 4.70 & 2.00  & 2.27& 0.00\\
\hline
g:CO WDs/He stars (all CE) & 13.35 & 0.20 & 24.24  & 11.87  & 13.46\\
h:CO WDs/He stars (Merger with $M_{\rm CO WD}\geq0.9$) & 1.61 & 0.00 & 1.20 & 0.10 & 0.69 \\
i:CO WDs/He stars (Merger with $M_{\rm CO WD}\geq1.0$) & 0.61 & 0.00 & 0.52 & 0.03 & 0.50  \\
\hline
\end{tabular}\\
Ranges of delay times (DT in Gyr) under five different models
\begin{tabular}{lccccccc}
\hline
Same as above & DT1 & DT2& DT3 & DT4 & DT5 \\

\hline
 a & 0.03 - 13.7 & 0.03 - 13.7 & 0.03 - 13.7   & 0.03 - 13.7 & 0.03 - 13.7 \\
 b & 0.07 - 0.71 & 0.07 - 0.71 & 0.05 - 1.61   & 0.05 - 0.71 & 0.00 - 0.00 \\
 c & 0.07 - 13.6 & 0.07 - 13.6 & 0.07 - 13.6   & 0.09 - 13.6 & 0.09 - 13.6 \\
 d & 0.08 - 13.6 & 0.08 - 13.6 & 0.08 - 13.6   & 0.08 - 13.6 & 0.08 - 13.6 \\
 e & 0.06 - 1.78 & 0.07 - 1.52 & 0.05 - 2.52   & 0.08 - 0.61 & 0.08 - 0.61 \\
 f & 0.07 - 1.78 & 0.07 - 1.52 & 0.08 - 2.13   & 0.08 - 0.62 & 0.00 - 0.00 \\
\hline
g  & 0.05 - 1.69 & 0.04 - 0.51 & 0.05 - 1.80   & 0.04 - 1.58 & 0.04 - 1.58 \\
h  & 0.05 - 0.70 & 0.00 - 0.00 & 0.05 - 0.52   & 0.04 - 1.07 & 0.04 - 1.07 \\
i  & 0.04 - 0.50 & 0.00 - 0.00 & 0.05 - 0.42   & 0.04 - 1.06 & 0.04 - 1.06 \\
\hline
\end{tabular}
\end{center}
Note: Normal stars (only stars with masses $< 8M_\odot$ in this table) mean sub-giant or giant branch stars with hydrogen-rich envelopes; He-rich stars are the (sub)giant helium (no hydrogen) stars.
The merger con1 means that the merger with the condition of $M_{\rm CO WD}\geq0.9$ and $M_{2\rm C}\geq0.6$, and  merger con2 is the merger with condition of $M_{\rm CO WD}\geq1.0$ and $M_{2\rm C}\geq0.6$. $M_{2\rm C}$ is the core mass of the WD's companion star. Event numbers can be easily calculated with $R\times10^7$, and the birthrate can be calculated by star formation rate ($SFR$) and binary fraction ($f_{\rm bin}$), $SFR\times f_{\rm bin}\times R$. For mergers, the delay time is the time started from the evolution of MS-MS binary to just before the formation of WD CE event.
\end{table}

\clearpage

\begin{table}

\begin{center}
\caption{Rates of merger events between CO WDs and cores of massive stars ($M_2\geqslant8.0M_\odot$) inside the CE, and rates of CO WD - NS systems that survive from the CE phase under different models with different cases (in $10^{-5}$ ${M_\odot}^{-1}$)}

\begin{tabular}{lccc|cc}
 \hline\hline

Model&Merger ($M_{\rm WD} \geq 0.9$)& Merger ($M_{\rm WD} \geq 1.0$)& DT (in Myr)& COWD-NS \\
\hline
 mod1 &5.05 & 2.09 & 34 - 123 & 0.54  \\
 mod2 & 12.18 & 4.59 &34 - 123 & 0.0 \\
 mod3 & 8.94 & 2.24 &42 - 113  & 0.86  \\
 mod4 & 3.54 &  0.001 & 34 - 123   & 2.72 \\
 mod5 & 0.85 & 0.0  & 34 - 117   & 2.83  \\

\hline
\end{tabular}
\end{center}
Note: DT means delay time for mergers in the case of $M_{\rm WD} \geq 0.9$. DTs of merger inside the CE are all shorter with a massive star, and there is no big difference for mergers with $M_{\rm WD} \geq 1.0$, thus we just show the range of DTs for one case.
\end{table}

\clearpage


\begin{figure}
\centering
\includegraphics[totalheight=2.4in,width=2.9in]{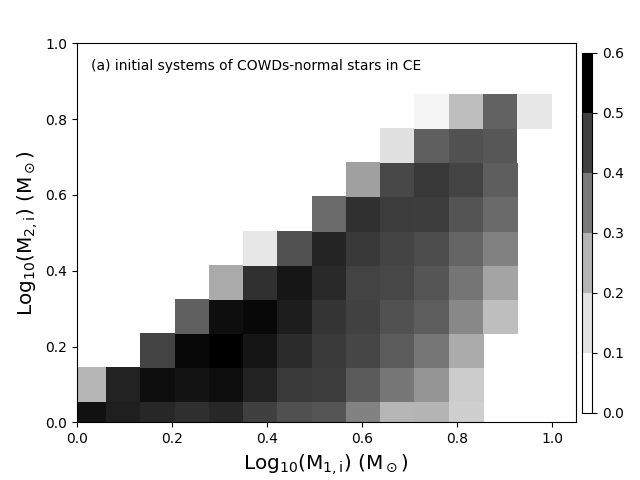}
\includegraphics[totalheight=2.4in,width=2.9in]{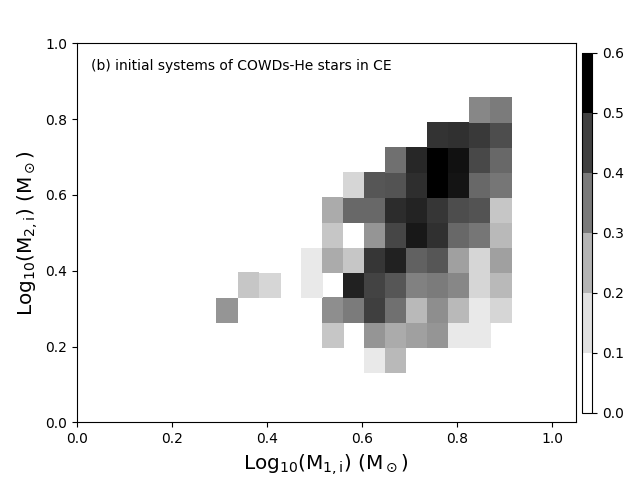}
\includegraphics[totalheight=2.4in,width=2.9in]{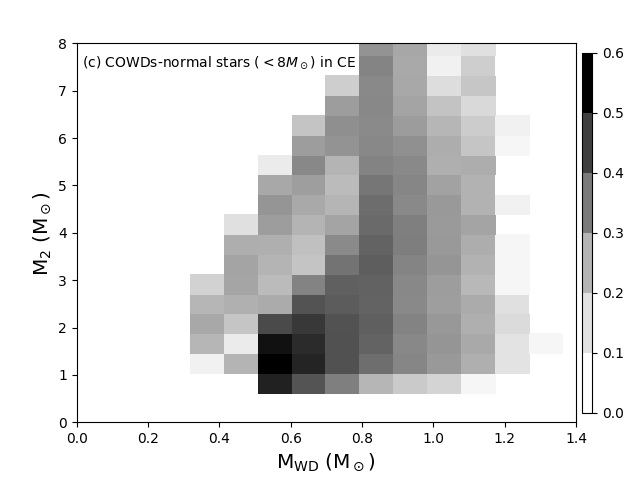}
\includegraphics[totalheight=2.4in,width=2.9in]{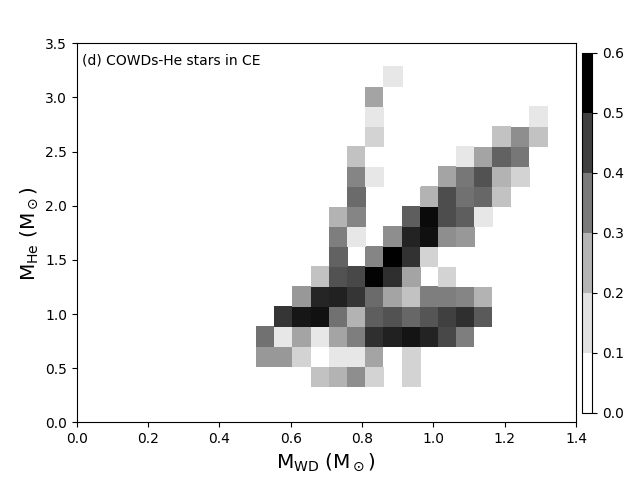}
\caption{The number density map from simulated results of model 1 (standard model). The numbers shown here are out of the all $10^7$ binaries in a simulation: the event counts of each cell is used to weight the density of each grid cell with a chosen grid cell size. For example, with the total event number of 800 resulted from the all $10^7$ binaries, it is weighted for each cell of the parameter space. (a) and (b) show the initial mass relation of primary MS ($M_{\rm 1,i}$) and secondary MS stars ($M_{\rm 2,i}$) in primordial binaries which form CO WD + normal star binaries (upper left) and CO WD + He star systems (upper right) which experience the CE. Mass relations at the beginning of the CE between the CO WD and its companion star for all CO WD + normal star systems and CO WD + He star binaries are given in (c) and (d), respectively. }
\label{fig:1}
\end{figure}

\clearpage

\begin{figure}
\centering
\includegraphics[totalheight=2.4in,width=2.9in]{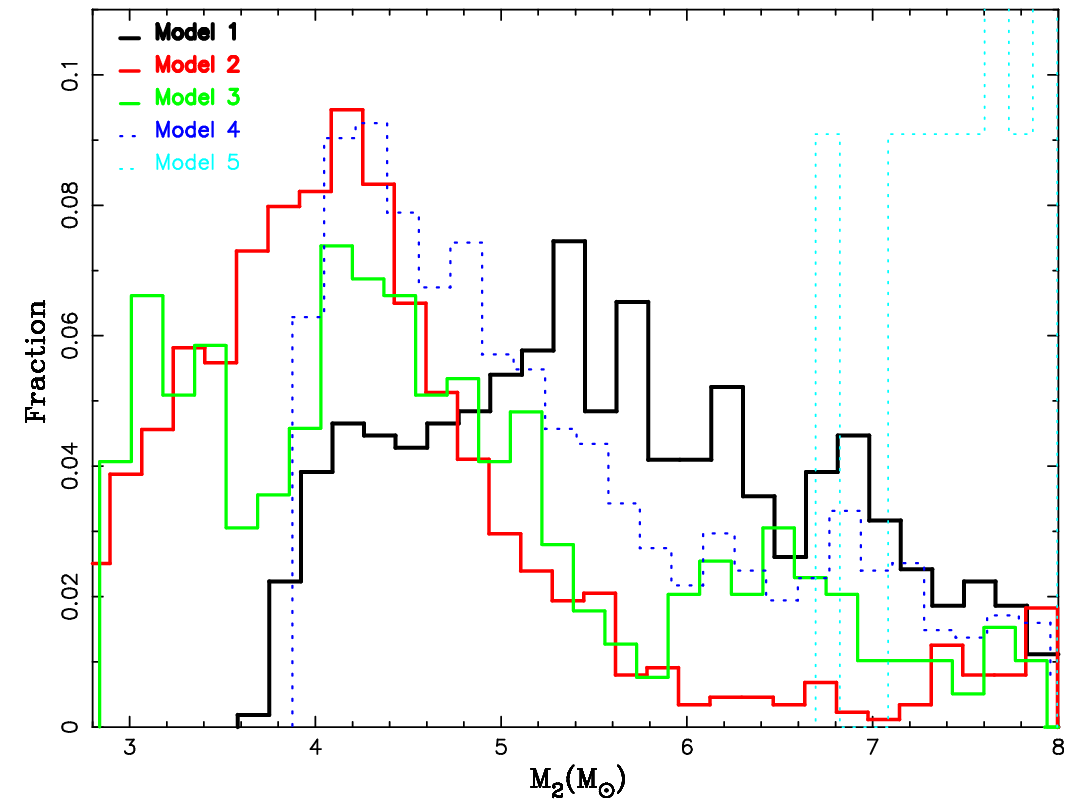}
\includegraphics[totalheight=2.4in,width=2.9in]{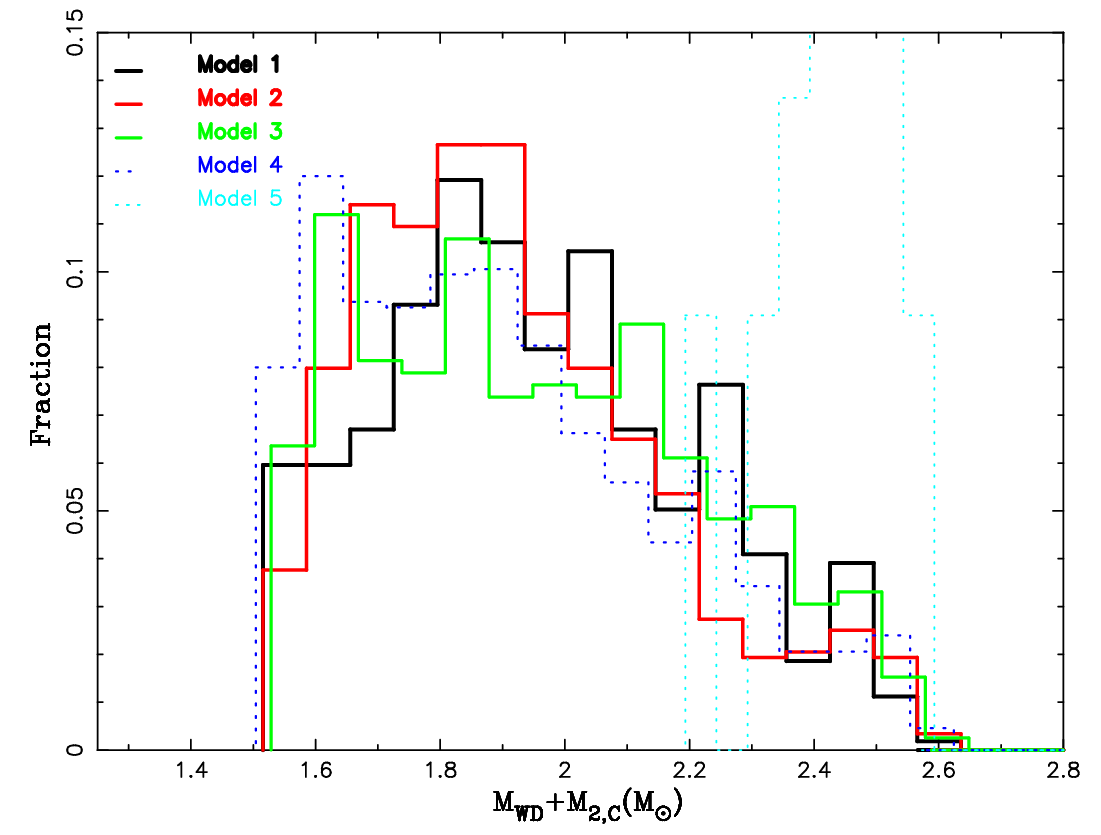}
\includegraphics[totalheight=2.4in,width=2.9in]{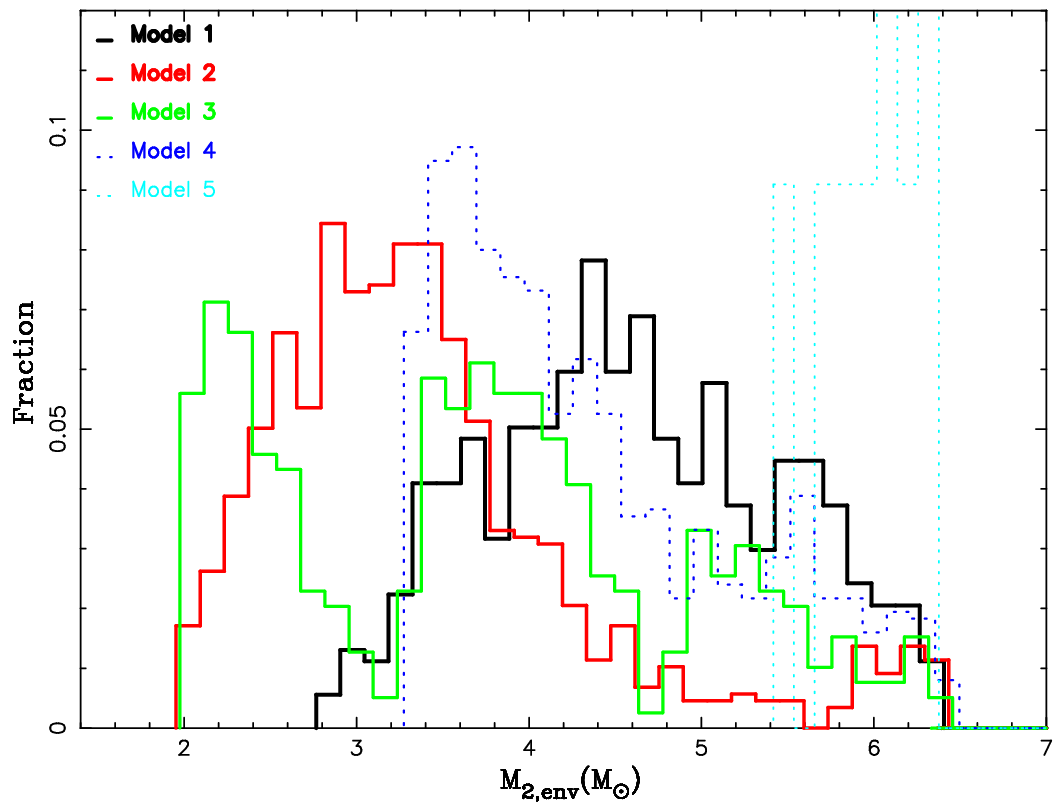}
\includegraphics[totalheight=2.4in,width=2.9in]{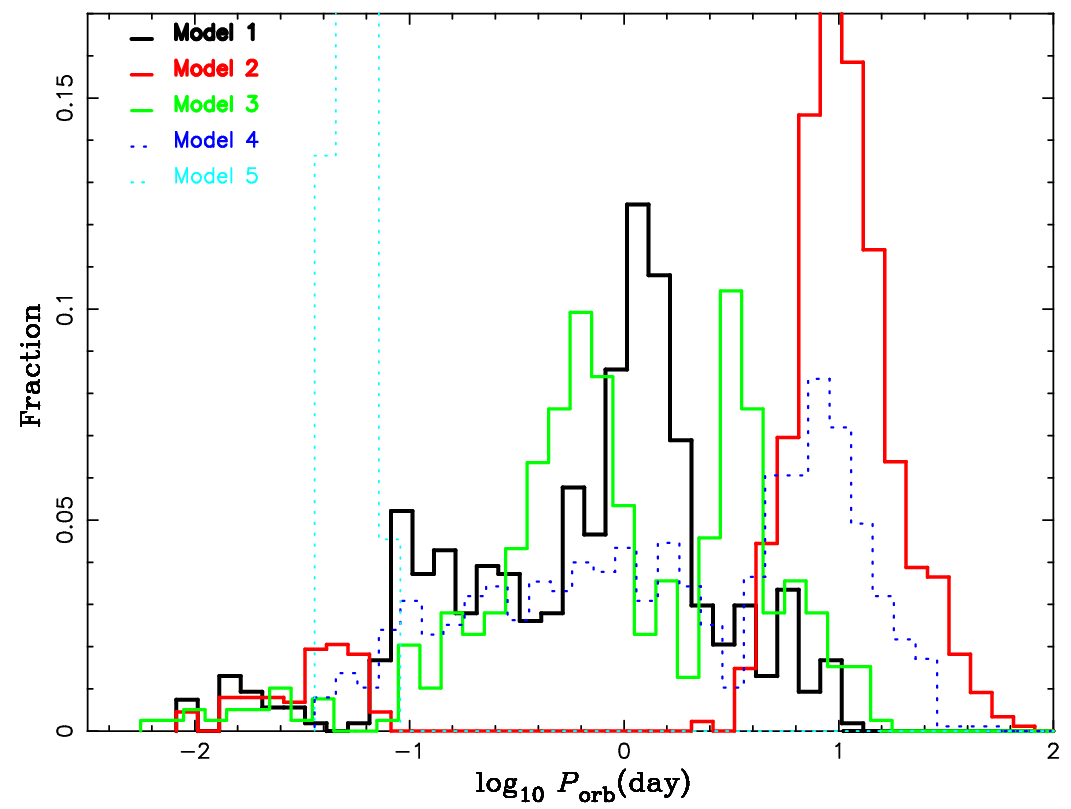}
\caption{The distributions of properties of merger systems between the CO WDs and He cores of the intermediate normal stars under the condition of $M_{\rm 2,C}\geq0.6$ and $M_{\rm WD}\geq0.9$ just before entering the CE, and simulated results from five models are given. The upper left panel shows the secondary mass distributions, the upper right panel gives the distributions of CO WD $+$ cores of the secondaries, the lower left one is for the distributions of secondaries' H-rich envelopes, and the lower right panel shows the orbital period distributions.}
\label{fig:1}
\end{figure}

\clearpage

\begin{figure}
\centering
\includegraphics[totalheight=4.0in,width=4.5in]{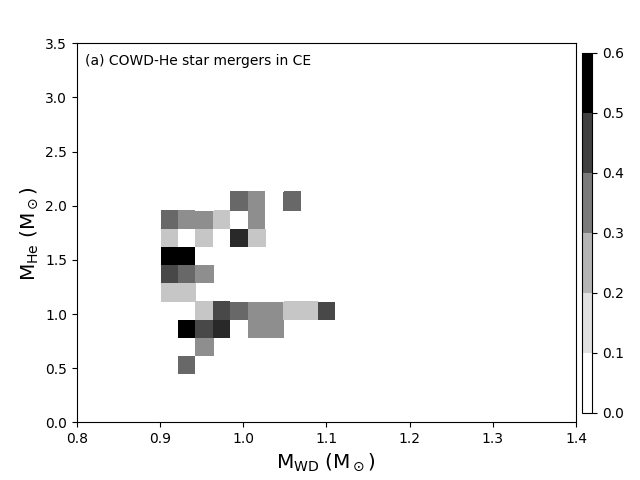}
\includegraphics[totalheight=4.0in,width=4.5in]{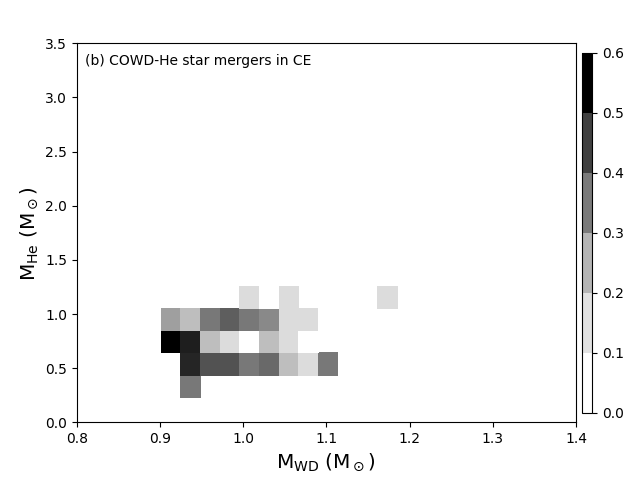}
\caption{The distributions from simulated results of model 1 (standard model) for the merger case of the CO WD and core of the He stars under under the condition of $M_{\rm 2,C}\geq0.6$ and $M_{\rm WD}\geq0.9$ at the onset of CE. (a) shows the mass relation of the CO WD ($M_{\rm WD}$) and secondary He stars ($M_{\rm He}$). Mass relations between the CO WD and core of the He star for the merger case are given in (b).}
\label{fig:1}
\end{figure}

\clearpage

\begin{figure}
\centering
\includegraphics[totalheight=2.4in,width=2.9in]{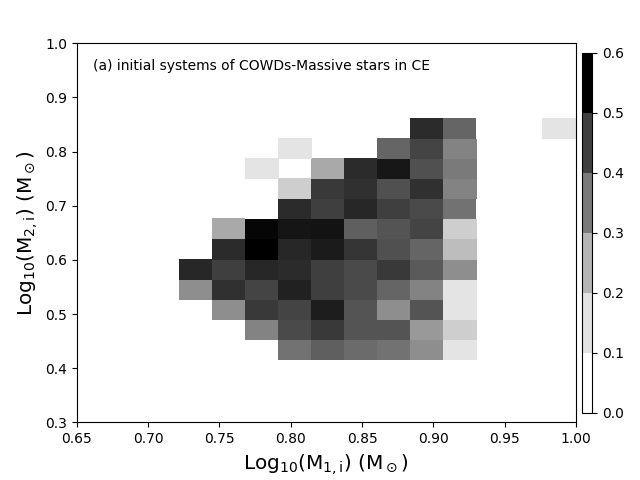}
\includegraphics[totalheight=2.4in,width=2.9in]{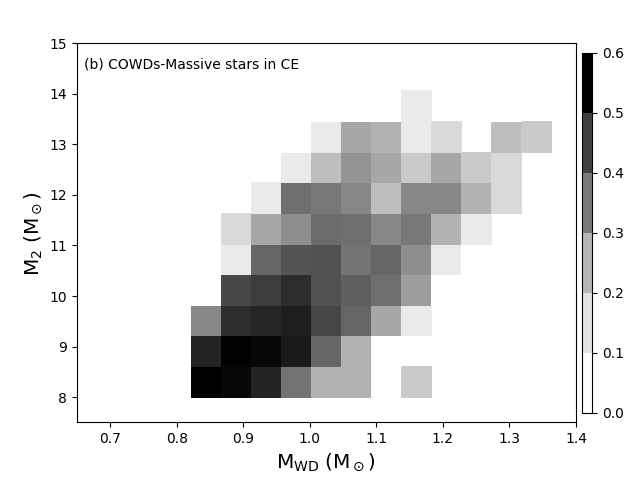}
\includegraphics[totalheight=2.4in,width=2.9in]{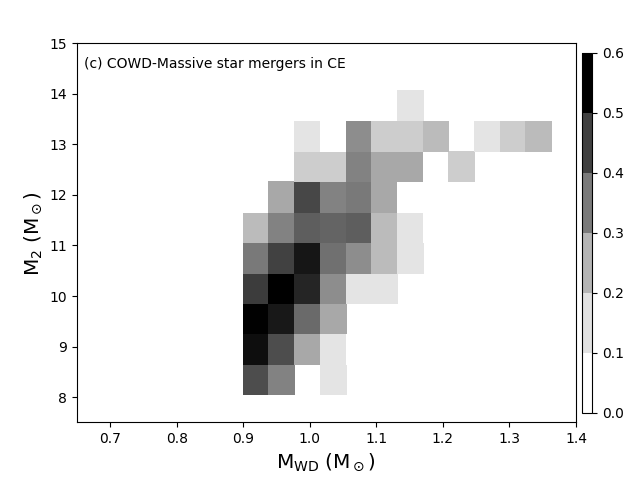}
\includegraphics[totalheight=2.4in,width=2.9in]{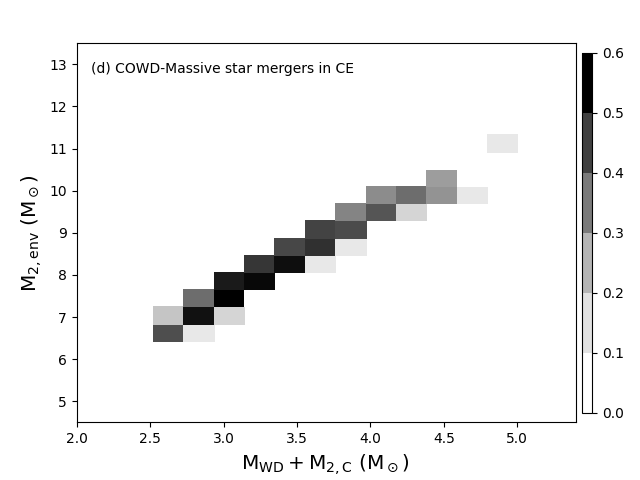}
\caption{The number density distributions from simulated results of model 1 (standard model). (a) shows the initial mass relation of primary MS ($M_{\rm 1,i}$) and secondary MS stars ($M_{\rm 2,i}$) in primordial binaries which form CO WD + massive star binaries (upper left) which experience the CE. Mass relations at the beginning of the CE between the CO WD and its companion star for all CO WD + massive star systems with CE are given in (b). (c) gives the mass relation between the CO WDs and massive star companions at the onset of the CE for merger case. Mass relation of the CO WD + core of the massive star and the H-rich envelope of the massive star for the binaries which will have the core mergers inside the CE is shown in (d). Note that `in CE' in the insets means that merger takes place during the CE evolution.}
\label{fig:1}
\end{figure}

\clearpage

\begin{figure}
\centering
\includegraphics[totalheight=4.5in,width=4.8in]{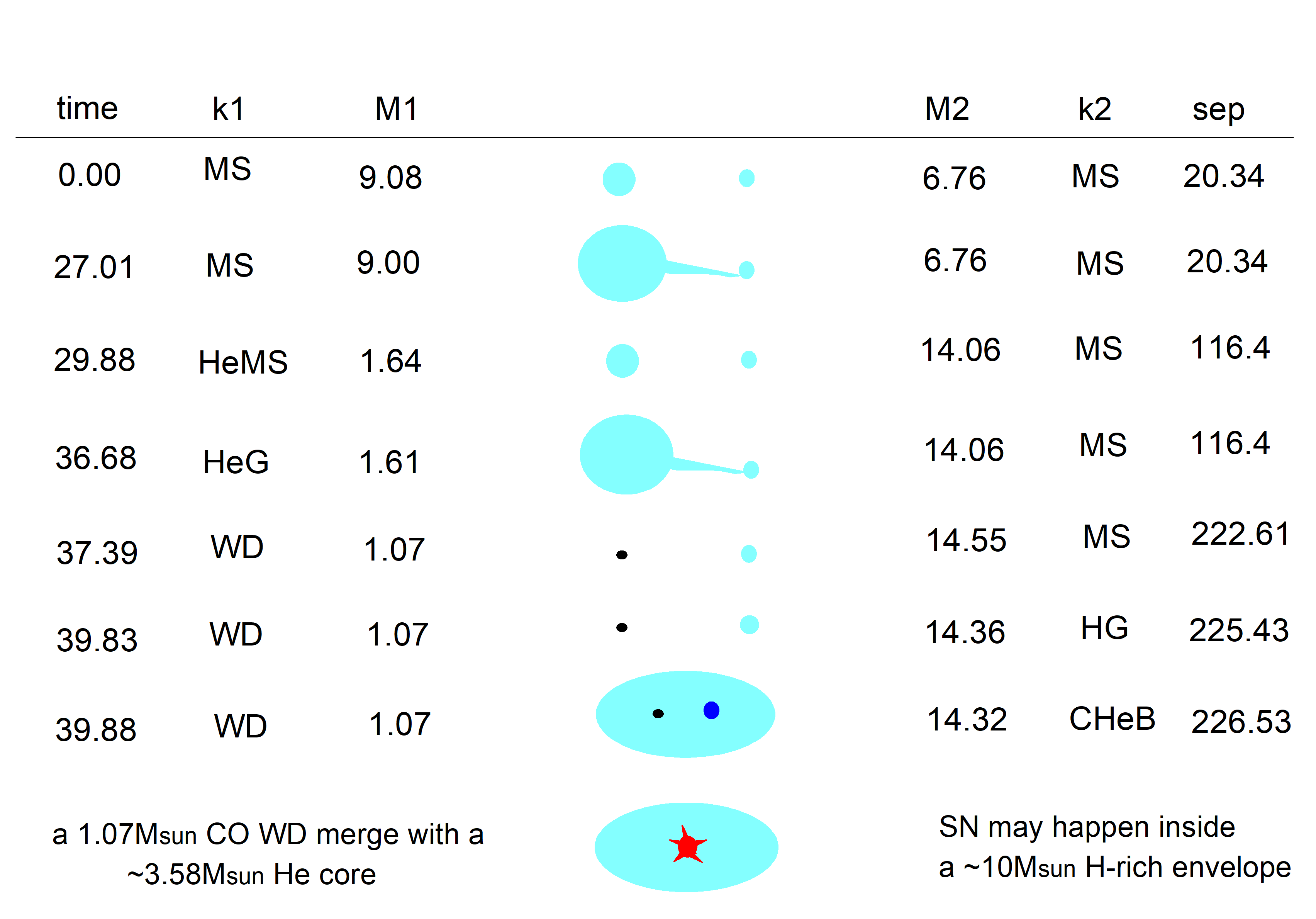}
\caption{An example of the evolution of a system to the channel for the merger between
a CO WD and He core of a massive star based on model 1. The primary (mass is M1) experiences two phases of mass transfer before forming
the CO WD. Afterwards, the secondary (mass is M2) fills its
Roche lobe at the core helium burning stage of the evolution (CHeB), and its envelope engulfs the CO WD and its core due to the unstable mass transfer.
The abbreviations of the stellar types (k1 and k2) are defined in the text. The evolution time is in Myr, and separation (sep) between the two stars is in $R_\odot$.}
\label{fig:1}
\end{figure}

\clearpage

\begin{figure}
\centering
\includegraphics[totalheight=4.5in,width=6.0in]{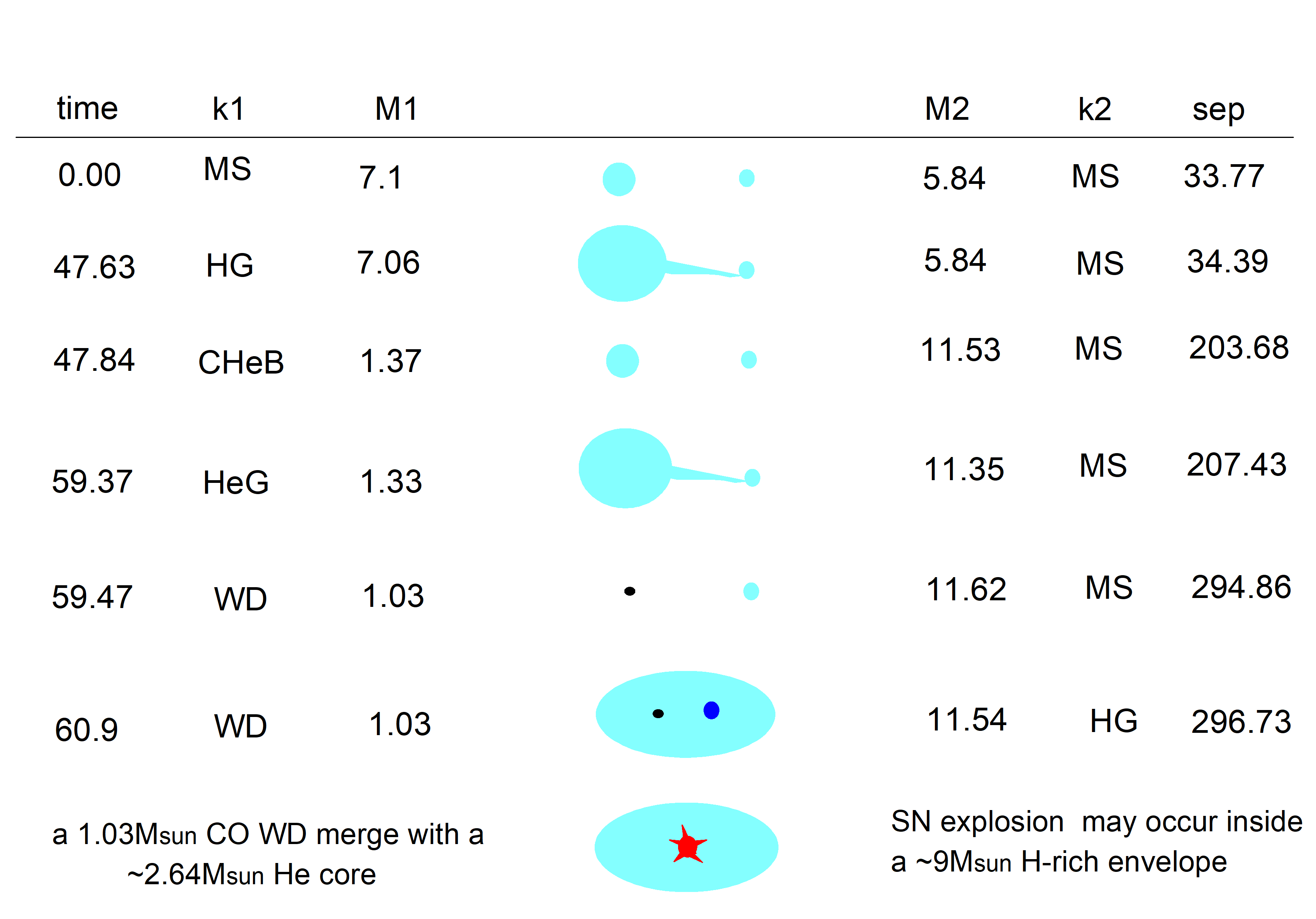}
\caption{Another example of the merger between a CO WD and He core of a massive star inside the CE based on model 1. The evolution is very similar to that of Fig. 1, and the main difference compared to the first example is that the CE event happens when the secondary becomes the HG star, and the evolution time is longer due to the different initial conditions.}
\label{fig:1}
\end{figure}

\clearpage

\begin{figure}
\centering
\includegraphics[totalheight=2.4in,width=2.9in]{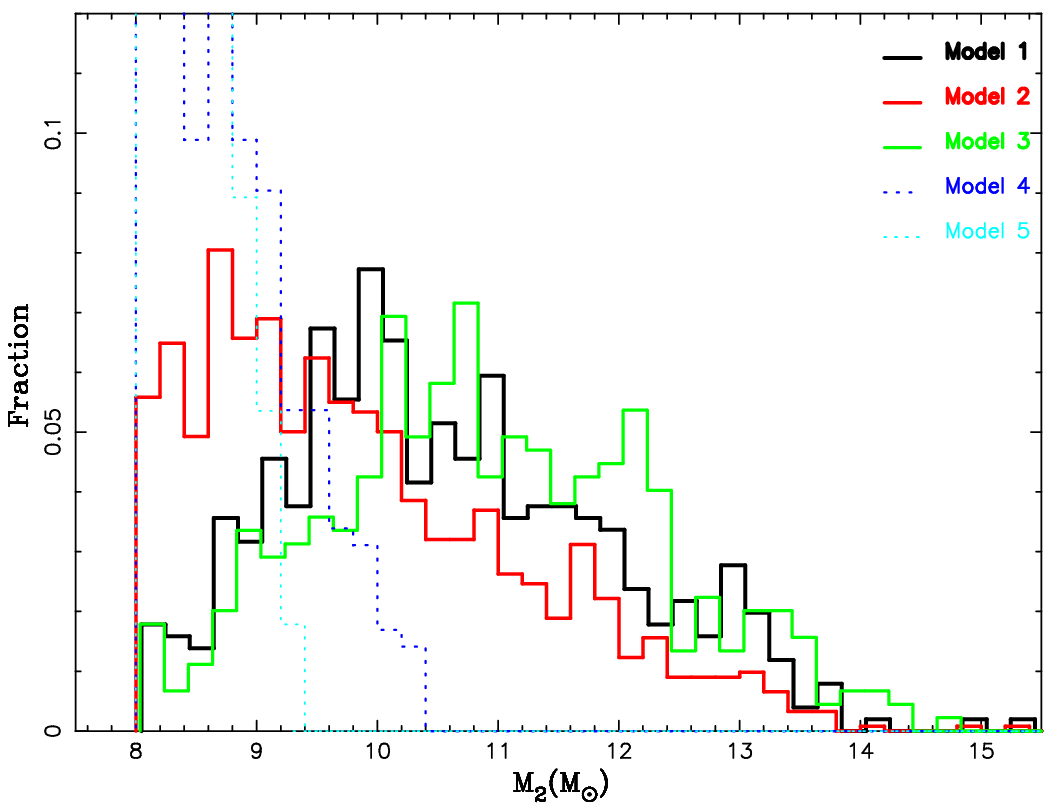}
\includegraphics[totalheight=2.4in,width=2.9in]{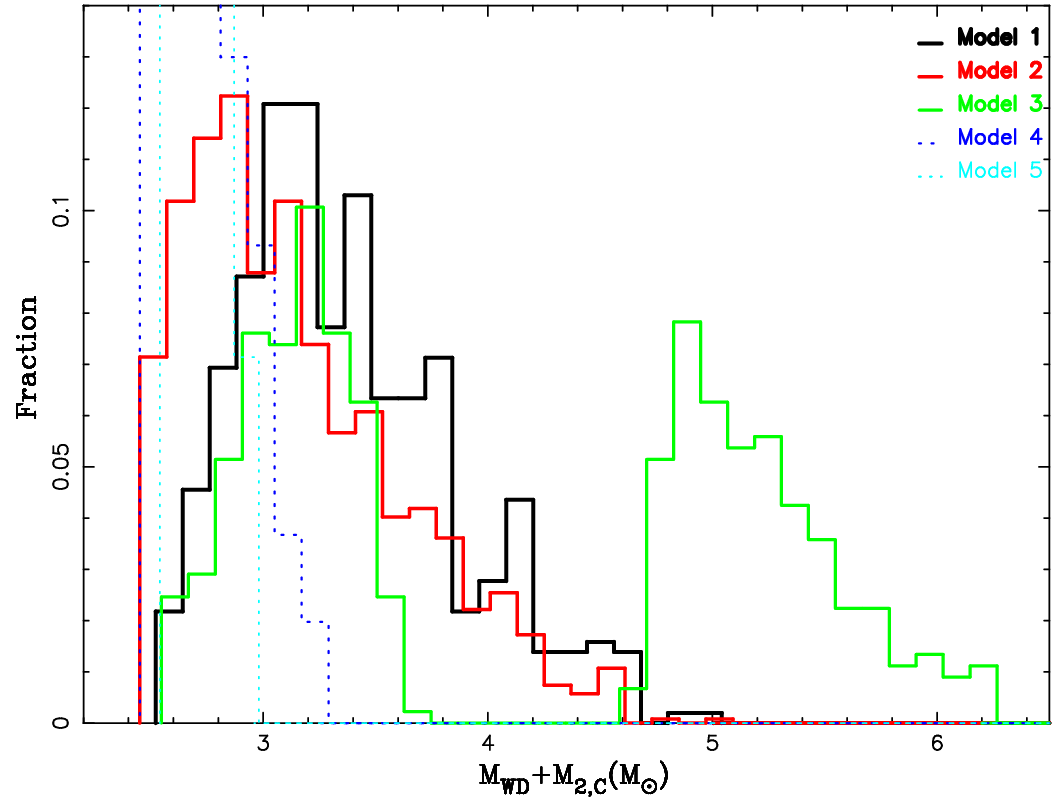}
\includegraphics[totalheight=2.4in,width=2.9in]{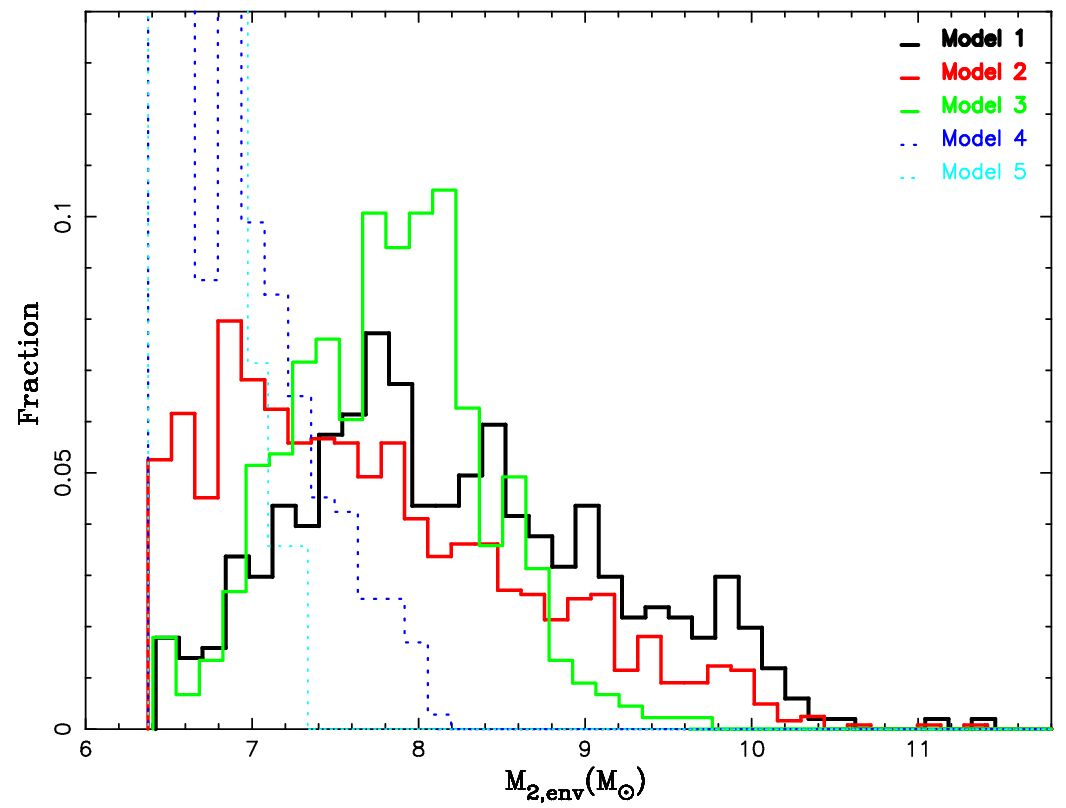}
\includegraphics[totalheight=2.4in,width=2.9in]{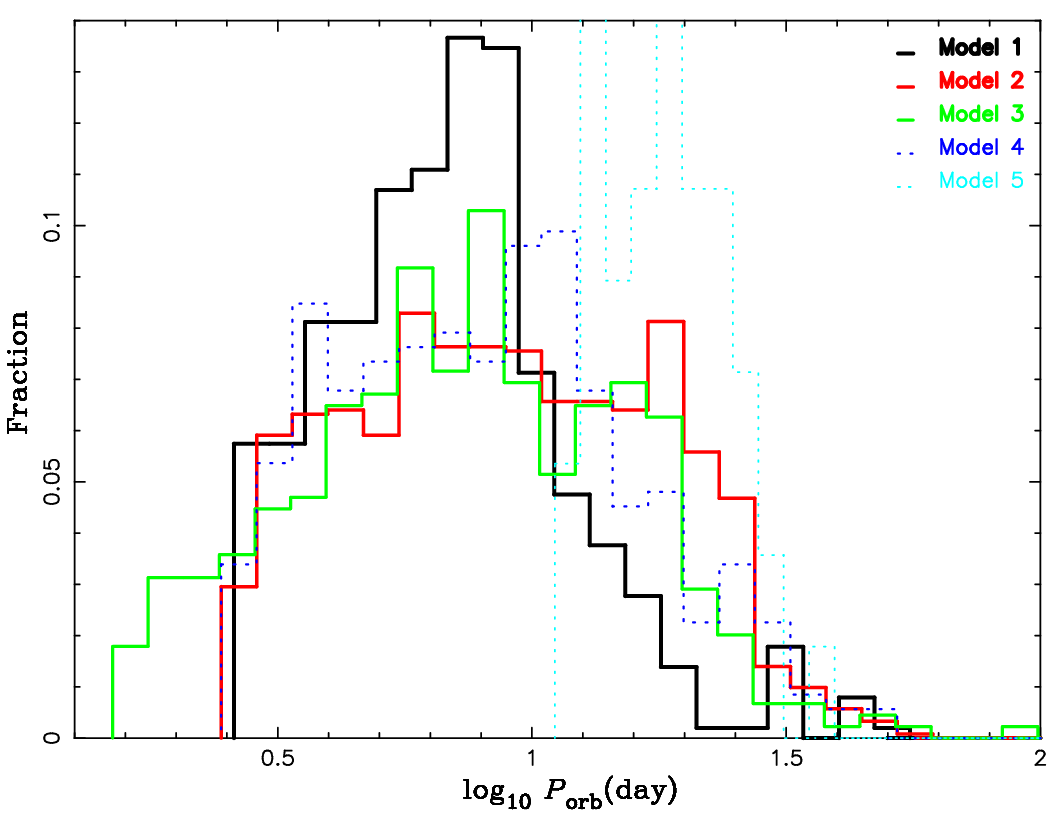}
\caption{The distributions of properties of merger systems between the CO WDs and He cores of massive stars under the condition of $M_{\rm WD}\geq0.9$ at the beginning of the CE, and simulated results from five models are shown. The upper left panel shows the secondary mass distributions, the upper right panel gives the distributions of CO WD $+$ cores of the secondaries, the lower left one is for the distributions of secondaries' H-rich envelopes, and the lower right panel shows the orbital period distributions.}
\label{fig:1}
\end{figure}

\clearpage

\begin{figure}
\centering
\includegraphics[totalheight=3.8in,width=4.4in]{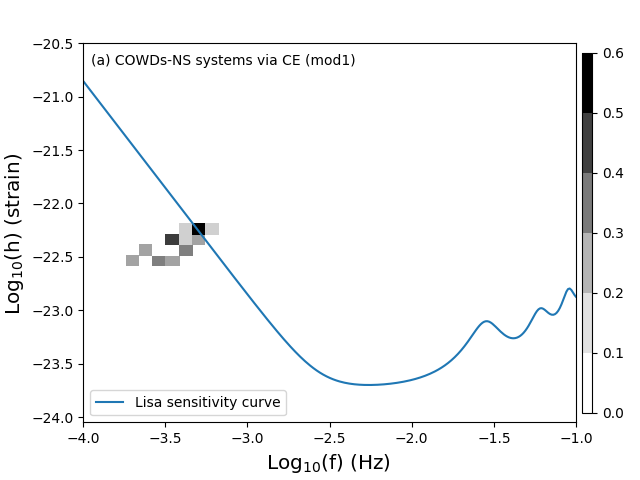}
\includegraphics[totalheight=3.8in,width=4.4in]{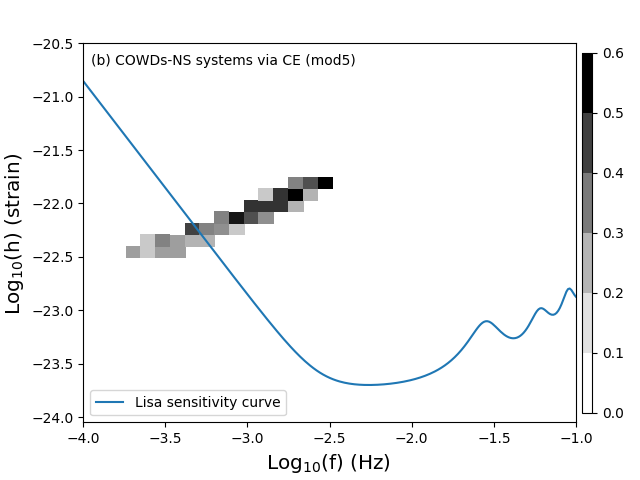}
\caption{The distribution of GW-sources from CO WDs - NSs (evolved through the CE described in text) mergers in the strain-frequency space. The sources are assumed to be at distances of 10 kpc.
The sensitivity curve of LISA is shown with the solid blue line. Gray scale is CO WD - NS sources in our model 1 (a) and model 5 (b) simulations.}
\label{fig:1}
\end{figure}

\clearpage



\end{document}